\documentclass[12pt]{iopart}
\usepackage{iopams}
\usepackage{amssymb}
\usepackage{amsthm}
\usepackage{latexsym}
\usepackage[dvips]{epsfig}
\usepackage{mathrsfs}

\theoremstyle{plain}
\newtheorem{proposition}{Proposition}
\newtheorem{remark}{Remark}
\newtheorem{lemma}{Lemma}
\newtheorem{theorem}{Theorem}

\newtheorem{definition}{Definition}

\newcommand{\DAl}{\raise -0.4mm \hbox{\Large$\Box$}}
\newcommand{\DAlhat}{{\raise -0.4mm \hbox{\Large$\Box$}}\hspace{-10.7pt}{\raise -1mm \hbox{\Large$\hat{\phantom{\Psi}}$}}}

\newcommand{\updn}[3]{#1^{#2}_{\phantom{#2}#3}}
\newcommand{\dnup}[3]{#1_{#2}^{\phantom{#2}#3}}

\eqnobysec

\begin{document}

\title[Kerr initial data]{Kerr initial data}

\author{Alfonso Garc\'{\i}a-Parrado G\'omez-Lobo}

\address{Matematiska institutionen, Link\"opings Universitet,\\
SE-581 83 Link\"oping, Sweden.}
\ead{algar@mai.liu.se}

\author{Juan A. Valiente Kroon}

\address{School of Mathematical Sciences, Queen Mary, University of London, \\
Mile End Road, London E1 4NS, UK.}
\ead{j.a.valiente-kroon@qmul.ac.uk}

\begin{abstract}
Exploiting a 3+1 analysis of the Mars-Simon tensor,
  conditions on a vacuum initial data set ensuring that its
  development is isometric to a subset of the Kerr spacetime are
  found. These conditions are expressed in terms of the vanishing of a
  positive scalar function defined on the initial data hypersurface.
  Applications of this result are discussed.
\end{abstract}

\submitto{\CQG}
%\maketitle

\section{Introduction}
The Kerr spacetime is one of the most important exact solutions to the
Einstein vacuum equations. Its relevance stems from the
uniqueness theorems for black holes which state that under rather
general conditions the Kerr spacetime is the only asymptotically
flat, stationary, vacuum black hole solution ---see e.g. the introduction of
\cite{IONESCU} for a critical review of the issue of black hole
uniqueness and the involved assumptions. Although, the Kerr spacetime
is very well understood from a spacetime perspective, the same cannot
be said if one adopts a 3+1 point of view ---which would be the case
if one tries to numerically calculate the spacetime from some Cauchy
initial data.

As soon as one moves away from a 3+1 gauge which is adapted to the
stationary and axial symmetries ---which can occur in some
applications, in particular in numerical ones--- an analysis of the
Kerr spacetime and initial data sets thereof becomes very complicated.
The explicit nature of the Kerr solution makes it tempting to perform
detailed calculations in order to, say, verify a particular property
of the spacetime. This approach usually leads to very lengthy
expressions which can be very hard to analyse.  In a sense one could
say that exact solutions contain too much information. In this case it
can be more convenient to adopt more abstract approaches and,
accordingly, it may prove useful to have at hand a characterization of
Kerr initial data.

The question of providing an invariant characterization of initial
data sets for the Schwarzschild spacetime has been addressed in
\cite{VALIENTE,GARVAL}. In particular, the analysis of \cite{GARVAL}
provides an algorithmic characterization of Schwarzschild data. That is, a
procedure is provided to verify whether a given initial data set for
the Einstein field equations will render a development which is
isometric to a portion of the Schwarzschild spacetime.

One of the most
important algebraic properties of the Kerr spacetime is that its Weyl
tensor is of Petrov type D. The close relation between vacuum
spacetimes with a Weyl tensor of Petrov type D and Killing spinors has
been exploited in \cite{GARVAL08} to provide a characterization of
initial data sets whose developments will be of Petrov type D. This
characterization relies on one being able to decide whether a set of
overdetermined partial differential equations has  solutions for a
given initial data set. Accordingly, such a characterization is not
algorithmic. Although not explicitly stated in \cite{GARVAL08}, from
that analysis it should be possible to obtain a characterization of
Kerr initial data by adding some global conditions.

The characterization of initial data sets discussed in
\cite{VALIENTE,GARVAL,GARVAL08} has followed the general strategy of
starting from a given tensorial (respectively, spinorial)
\emph{spacetime} characterization of the class of spacetimes under
consideration. Necessary conditions on the initial data set are obtained by
performing a 3+1 decomposition of the spacetime characterization.
Given a set of necessary conditions on the initial data, it is then
natural to address the question of sufficiency. This is, usually, the
most challenging part of the procedure as one has to discuss the
evolution of complicated tensorial objects. The idea behind this is to show that if the
necessary conditions are satisfied on some open subset of the initial
hypersurface, then one can ---possibly, under some additional
assumptions--- recover the spacetime characterization on the
development of the open subset on the initial hypersurface from which
one started.

In this article a particular characterization of Kerr initial data is
addressed. Our starting point is a certain spacetime characterization
of the Kerr spacetime given in \cite{MARS-1,MARS-2}. This
characterization was developed with the aim of providing an
alternative way of proving the uniqueness of Kerr spacetime among the
class of stationary, asymptotically flat black holes. This expectation
has been recently fulfilled in \cite{IONESCU}, where a proof of
the uniqueness of Kerr which does not make assumptions on the
analyticity of the metric has been given. At the heart of the spacetime
characterization given in \cite{MARS-1,MARS-2} ---cfr. theorem
\ref{kerr-characterization}--- is a certain tensor, \emph{the Mars-Simon
  tensor}, whose construction requires the existence of a timelike
Killing vector. The Mars-Simon tensor is a spacetime version of the
\emph{Simon tensor}, a tensor, defined in the quotient manifold of a
stationary spacetime, which characterizes the Kerr spacetime ---see
\cite{SIMON}.

Following the general strategy for the construction of
characterizations of initial data sets out of spacetime
characterizations, necessary conditions for Kerr initial data are deduced
from a 3+1 splitting of the Mars-Simon tensor. Accordingly, one
assumes that the spacetime  one is working with has a timelike Killing
vector. This requirement can be encoded in the initial data by
requiring that the data has a suitable \emph{Killing initial data}
(KID). The Mars-Simon tensor has the same symmetries as the Weyl
tensor, and hence its 3+1 decomposition can be given in terms of its
\emph{electric} and \emph{magnetic} parts. In order to discuss the
propagation of the Mars-Simon tensor we make use of a framework for
discussing the propagation tensorial fields using
\emph{superenergy-type} tensors ---this framework has been discussed
in e.g. \cite{BERGQVIST}. It should be pointed out that the
characterization discussed in this article is not algorithmic. That
is, like the one for type D initial data discussed in \cite{GARVAL08}
it depends on being able to decide whether a certain overdetermined
system of partial differential equations admits a solution.

\bigskip The article is structured as follows: in section
\ref{preliminaries} our main conventions are fixed and relevant
aspects of the 3+1 formalism are discussed. Section \ref{weyl:c}
discusses the properties and causal propagation of Weyl candidates
---i.e. tensors with the same symmetries of the Weyl tensor. Section
\ref{section:mstensor} is concerned with the properties of the
Mars-Simon tensor. Section \ref{cp-ms} discusses the causal
propagation of the Mars-Simon tensor. Section
\ref{section:application} applies the previous discussion to the
construction of a characterization of Kerr initial data. Our main
result is provided in theorem \ref{theorem:Kerr_data}. In section
\ref{Schwarzschild} we particularize to the case of the Schwarzschild
spacetime, where, with the aim of the results of \cite{GARVAL} it is
possible to obtain an initial data characterization which is
algorithmic. Some concluding remarks are provided in section
\ref{conclusions}. Finally, some technical details, too lengthy to be
presented in the main text are presented in the appendix A.

All the tensor computations of this paper have been performed with the
software {\em xAct} \cite{XACT}. {\em xAct} is a suite of MATHEMATICA
packages which has among its many features the capability to
efficiently canonicalize tensor expressions by the use of powerful
algorithms based on permutation group theory \cite{XPERM}.  Currently
no other software, either free or commercial, is capable to handle the
tensor computations needed in this paper.

\section{Preliminaries} \label{preliminaries} Let
$(\mathcal{M},g_{\mu\nu})$ denote a smooth orientable spacetime. The
following conventions will be used: plain Greek letters
$\alpha,\beta,\gamma,\dots$ denote abstract indices and boldface Latin
characters $\boldsymbol{a}, \boldsymbol{b}, \boldsymbol{c},\dots$ will
be used for component indices. The signature of the metric tensor
$g_{\mu\nu}$ will be taken to be $(-,+,+,+)$, while
$R_{\mu\nu\alpha\rho}$,
$R_{\mu\nu}=R^{\alpha}_{\phantom{\alpha}\mu\alpha\nu}$ and
$W_{\alpha\beta\mu\nu}$ denote, respectively the Riemann, Ricci and
Weyl tensors of $g_{\mu\nu}$. The tensor $\eta_{\alpha\beta\sigma\nu}$
is the volume element which is used to define the Hodge dual of any
antisymmetric tensor ---denoted by attaching a star $*$ to the tensor
symbol. Sometimes we will need to work with complex tensors in which
case the complex conjugation of a tensor is denoted by an overbar. The
operator $\mathcal{L}_{\vec{u}}$ symbolizes the Lie derivative with respect
to the vector field $u^\mu$.

\subsection{The orthogonal splitting}\label{section:osplitting}
Let $n^\mu$ be a unit timelike vector, $n^\mu n_\mu=-1$ defined on
$\mathcal{M}$. Then any tensor or tensorial expression can be
decomposed with respect to $n^\mu$ and the way to achieve it is the
essence of the orthogonal splitting (also known as 3+1 formalism)
which is described in many places of the literature ---see e.g.
\cite{ELLIS,GOURG}). We review the parts of this formalism needed
in this work. The spatial metric is defined by $h_{\mu\nu}\equiv
g_{\mu\nu}+n_\mu n_\nu$ and it has the algebraic properties
$h^\mu_{\phantom{\mu}\mu}=3$,
$h_\mu^{\phantom{\mu}\sigma}h_{\sigma\nu}=h_{\mu\nu}$. We shall call a
covariant tensor $T_{\alpha_1\dots \alpha_m}$ {\em spatial} with
respect to $h_{\mu\nu}$ if it is invariant under
$h^\mu_{\phantom{\mu}\nu}$ i.e. if
\begin{equation}
h^{\alpha_1}_{\phantom{\alpha_1}\beta_1}\cdots h^{\alpha_m}_{\phantom{\alpha_m} \beta_m}T_{\alpha_1\cdots \alpha_m}=T_{\beta_1\cdots\beta_m},
\end{equation}
with the obvious generalization for any mixed tensor. This property
implies that the inner contraction of $n^\mu$ with $T_{\alpha_1\dots
  \alpha_m}$ (taken on any index) vanishes.  The orthogonal splitting
of a tensor expression consists in writing it as a sum of terms which
are tensor products of the unit normal and spatial tensors of lesser
degree ---or the same degree in which case the unit normal is absent.

In order to find the orthogonal splitting of expressions containing
covariant derivatives we need to introduce the {\em spatial
  derivative} $D_\mu$ which is an operator whose action on any tensor
field $\updn{T}{\alpha_1\dots\alpha_p}{\beta_1\dots\beta_q}$,
$p,q\in\mathbb{N}$ is given by
\begin{equation}
\label{spatialcd}
D_\mu \updn{T}{\alpha_1\dots\alpha_p}{\beta_1\dots\beta_q}\equiv
h^{\alpha_1}_{\phantom{\alpha_1}\rho_1}\dots h^{\alpha_p}_{\phantom{\alpha_p} \rho_p}h^{\sigma_1}
_{\phantom{\sigma_1}\beta_1}\dots
h^{\sigma_q}_{\phantom{\sigma_q}\beta_q}h^{\lambda}_{\phantom{\lambda}\mu}
\nabla_\lambda
\updn{T}{\rho_1\dots\rho_p}{\sigma_1\dots\sigma_q}.
\end{equation}
From equation (\ref{spatialcd}) it is clear that $D_\mu
\updn{T}{\alpha_1\dots\alpha_p}{\beta_1\dots\beta_q}$ is spatial.

The results just described hold for an arbitrary unit timelike vector
$n^\mu$ but in our framework we only need to consider integrable
timelike vectors which are characterized by the condition
$n_{[\mu}\nabla_{\nu}n_{\sigma]}=0$ (Frobenius condition).
In this case there exists a
foliation of $\mathcal{M}$ such that the vector field $n^\mu$ is
orthogonal to the leaves of the foliation. We shall denote by
$\{\Sigma_t\}$, $t\in I\subset\mathbb{R}$, the family of leaves of
this foliation and $\Sigma_0$ is called the {\em initial data
hypersurface} ---it is  assumed that $0\in I$. The tensor $h_{\mu\nu}$ plays
the role of the {\em first fundamental form} for any of the leaves
while the symmetric tensor $K_{\mu\nu}$ defined by
\begin{equation}
K_{\mu\nu}\equiv-\frac{1}{2}\mathcal{L}_{\vec{n}}h_{\mu\nu},
\label{lie-h}
\end{equation}
can be identified with the {\em second fundamental form}.  Combining
the previous definition with the Frobenius condition we easily derive
\begin{equation}
\nabla_\mu n_\nu=-K_{\mu\nu}-n_\mu A_\nu,
\label{gradnormaltoextrinsic}
\end{equation}
where $A^\mu\equiv n^\rho\nabla_\rho n^\mu$ is the acceleration of
$n^\mu$.  By using these quantities one is in principle able to work
out the orthogonal splitting of any tensorial expression. We supply
below the explicit results of calculations which will be needed
repeatedly in the sequel.

\begin{itemize}

 \item Orthogonal splitting of the volume element:
\begin{equation}
\eta_{\alpha\beta\gamma\delta}=-n_\alpha\varepsilon_{\beta\gamma\delta}+n_\beta
\varepsilon_{\alpha\gamma\delta}
-n_\gamma\varepsilon_{\alpha\beta\delta}+n_\delta\varepsilon_{\alpha\beta\gamma}.
\label{decompose_eta}
\end{equation}
Here $\varepsilon_{\alpha\beta\gamma}\equiv n^\mu\eta_{\mu\alpha\beta\gamma}$ is the {\em spatial volume element} which is a fully antisymmetric spatial tensor.

\item Orthogonal splitting of an antisymmetric tensor. If $H_{\mu\nu}$ is an
arbitrary antisymmetric tensor then its orthogonal splitting takes the form
\begin{equation}
H_{\mu\nu}=2 P_{[\nu}n_{\mu]}-Q_\rho\varepsilon_{\mu\nu}^{\phantom{\mu\nu}\rho},
\end{equation}
where $P_\mu\equiv H_{\mu\nu}n^\nu$, $Q_{\mu}\equiv H^*_{\mu\nu}n^\nu$
are clearly spatial tensors. It is straightforward from the previous
equation to find an analogous formula for the orthogonal splitting of
$H^*_{\mu\nu}$.  If we work with the {\em self-dual} part of
$H_{\mu\nu}$, denoted by $\mathcal{H}_{\mu\nu}$ and defined by
\begin{equation}
\mathcal{H}_{\mu\nu}\equiv H_{\mu\nu}+\mbox{i}\ H^*_{\mu\nu},
\end{equation}
then its orthogonal splitting takes the form
\begin{equation}
\mathcal{H}_{\mu\nu}=\mbox{i}\ \varepsilon_{\rho\mu\nu}\mathcal{P}^\rho-2\mathcal{P}_{[\mu}n_{\nu]},\
\mathcal{P}_\mu\equiv P_\mu+\mbox{i}\ Q_\mu.
\label{decompose_H}
\end{equation}

\item
Orthogonal splitting of the Weyl tensor:
\begin{equation}
 W_{\mu\nu\lambda\rho}=2\left(l_{\mu[\lambda}E_{\rho]\nu}-
l_{\nu[\lambda}E_{\rho]\mu}-
n_{[\lambda}B_{\rho]\tau}\epsilon^\tau_{\phantom{\tau}\mu\nu}
-n_{[\mu}B_{\nu]\tau}\epsilon^\tau_{\phantom{\tau}\lambda\rho}
\right),\label{weyl-1}
\end{equation}
where
\begin{equation}
E_{\tau\rho}\equiv W_{\tau\nu\rho\lambda}n^\nu n^\lambda, \quad
B_{\tau\rho}\equiv W^*_{\tau\nu\rho\lambda}n^\nu n^\lambda,
\label{e-m}
\end{equation}
denote the Weyl tensor \emph{electric} and \emph{magnetic} parts
respectively, and $l_{\mu\nu}\equiv h_{\mu\nu}+n_{\mu}n_{\nu}$. The
tensors $E_{\mu\nu}$ and $B_{\mu\nu}$ are symmetric, traceless, and
spatial. Equation (\ref{weyl-1}) also holds for any rank-4 tensor
possessing the same algebraic properties as the Weyl tensor. Such tensors will be designated collectively as {\em Weyl candidates}
---see section \ref{weyl:c}.  We can also choose to work with the
self-dual Weyl tensor $\mathcal{W}_{\mu\nu\lambda\rho}$ which is given
by
\begin{equation}
\mathcal{W}_{\mu\nu\lambda\rho}\equiv
\frac{1}{2}(W_{\mu\nu\lambda\rho} + \mbox{i}\ W^*_{\mu\nu\lambda\rho}),
\end{equation}
Using (\ref{weyl-1}) and the analogous formula for the orthogonal
splitting of $W^*_{\mu\nu\lambda\rho}$ one can obtain an expression
for the orthogonal splitting of $\mathcal{W}_{\mu\nu\lambda\rho}$
\begin{equation}
\mathcal{W}_{\mu\nu\rho\lambda}=2(l_{\nu[\lambda}\mathcal{E}_{\rho]\mu}+
l_{\mu[\rho}\mathcal{E}_{\lambda]\nu}+\mbox{i}\ \varepsilon_{\rho\lambda\alpha}n_{[\mu}\mathcal{E}_{\nu]}^{\phantom{\nu]}\alpha}+
\mbox{i}\ \varepsilon_{\mu\nu\alpha}n_{[\rho}\mathcal{E}_{\lambda]}^{\phantom{\nu]}\alpha}),
\label{decompose-weyl}
\end{equation}
where
\begin{equation}
\mathcal{E}_{\mu\nu}\equiv\mathcal{W}_{\mu\rho\nu\lambda}n^\rho n^\lambda=
\frac{1}{2}(E_{\mu\nu}-\mbox{i}\ B_{\mu\nu}),\quad
\mathcal{E}_{(\mu\nu)}=
\mathcal{E}_{\mu\nu},\quad \mathcal{E}^\rho_{\phantom{\rho}\rho}=0.
\end{equation}
Again a similar formula to (\ref{decompose-weyl}) holds for any Weyl candidate.

\item Let $T^\mu$ be any spatial vector. Then the orthogonal splitting of
$\nabla_\mu T_\nu$ is
\begin{equation}
 \nabla_\mu T_\nu=D_\mu
T_\nu-n_\mu(K_{\rho\nu}T^\rho+\mathcal{L}_{\vec{n}}T_{\nu})
-T^\rho A_\rho n_\mu n_\nu-n_\nu K_{\rho\mu}T^\rho.
\label{decompose_cdT}
\end{equation}
Note that since $T_{\mu}$ is spatial then $\mathcal{L}_{\vec{n}}T_{\nu}$ is also spatial.
\end{itemize}

\subsection{The Cauchy problem}
We briefly review the standard formulation of the Cauchy problem for
the vacuum Einstein equations. In this formulation one considers a
3-dimensional connected Riemannian manifold $(\Sigma,h_{ij})$ ---we
use small plain Latin letters $i,j,k,\dots$ for the abstract indices
of tensors on this manifold--- and an isometric embedding
$\phi:\Sigma\longrightarrow\mathcal{M}$. The map $\phi$ is an
isometric embedding if $\partial_i \phi^\mu \partial_j \phi^\nu
g_{\mu\nu}=h_{ij}$ where $\partial_i \phi^\mu \partial_j \phi^\nu$
realises the pullback, $\phi^*$, of tensor fields from $\mathcal{M}$
to $\Sigma$.  The metric $h_{ij}$ defines a unique affine connection
$D_i$ (Levi-Civita connection) by means of the standard condition
\begin{equation}
D_jh_{ik}=0.
\end{equation}
The Riemann tensor of $D_i$ is $r_{ijkl}$ and from it we define its
Ricci tensor by $r_{ij}\equiv r^{l}_{\phantom{l}ilj}$ and its scalar
curvature $r\equiv r^i_{\phantom{i}i}$ ---in $\Sigma$ indices are
raised and lowered with $h_{ij}$ and its inverse $h^{ij}$.
\begin{theorem}
Let $(\Sigma,h_{ij})$ be a Riemannian manifold and suppose that
there exists a symmetric tensor field $K_{ij}$ on it which verifies
the conditions (vacuum constraints)
\numparts
\begin{eqnarray}
&& r+ K^2-K^{ij}K_{ij}=0, \label{Hamiltonian}\\
&& D^jK_{ij}-D_iK=0, \label{Momentum}
\end{eqnarray}
\endnumparts
where $K\equiv K^{i}_{\phantom{i}i}$. Provided that $h_{ij}$ and $K_{ij}$ are
suitably smooth there exists an isometric embedding $\phi$ of
$\Sigma$ into a globally hyperbolic, vacuum solution
$(\mathcal{N},g_{\mu\nu})$ of the Einstein field equations. The set
$(\Sigma,h_{ij}, K_{ij})$ is then called a vacuum initial data
set and the spacetime $(\mathcal{N},g_{\mu\nu})$ is the data
development. Furthermore the spacelike hypersurface
$\phi(\Sigma)$ is a Cauchy hypersurface in $\mathcal{N}$.
\label{vacuum-data}
\end{theorem}

A statement of this theorem containing precise regularity conditions on
$h_{ij}$, $K_{ij}$ is formulated in reference \cite{KLAINERMAN-NICOLO}.

Under the conditions of the theorem \ref{vacuum-data}, we may construct a
foliation of $\mathcal{N}$ with $n^\mu$ as the timelike unit vector
which is orthogonal to the leaves and, in the above notation, set
$\Sigma_0=\phi(\Sigma)$.  It is then clear that $\partial_i \phi^\mu
\partial_j \phi^\nu h_{\mu\nu}=h_{ij}$.  Other key properties are
\begin{equation}
\fl\partial_i \phi^\mu \partial_j \phi^\nu K_{\mu\nu}=K_{ij},\quad
\partial_i \phi^\mu \partial_{j_1}\phi^{\nu_1}\cdots \partial_{j_q}\phi^{\nu_q} (D_\mu T_{\nu_1\dots\nu_q})=
D_i(\partial_{\nu_1}\phi^{\nu_1} \cdots \partial_{\nu_q}\phi^{\nu_q} T_{\nu_1\dots\nu_q}).
\end{equation}
Using these properties and the Ricci identity for $n^\mu$ one can show
that
\numparts
\begin{eqnarray}
&& E_{ij}\equiv \partial_i \phi^\mu
\partial_j \phi^\nu E_{\mu\nu} =r_{ij}+K K_{ij}-K_{ik}K^k_{\phantom{k}j}
\label{electric_idata},
\\ && B_{ij}\equiv \partial_i \phi^\mu
\partial_j \phi^\nu B_{\mu\nu}=\epsilon_{(i}^{\phantom{(i}kl}D_{|k}K_{l|j)}\
\label{magnetic_idata}.
\end{eqnarray}
\endnumparts
Another property which is needed later on is the following one: for any spatial tensor $P_{\alpha_1\dots\alpha_p}$, $p\in\mathbb{N}$ with respect to $n^\mu$ we have the property
\begin{equation}
P_{\alpha_1\dots\alpha_p}|_{\phi(\Sigma)}=0\Longleftrightarrow
\phi^*(P_{\alpha_1\dots\alpha_p})=0.
\label{spatialp}
\end{equation}

\section{Weyl candidates}\label{weyl:c}

Fundamental for our discussion will be tensors of rank 4 having
the same algebraic properties as the Weyl tensor. More precisely, one defines a \emph{Weyl candidate} as any rank 4 tensor
$C_{\alpha\beta\gamma\delta}$ fulfilling the properties
\begin{equation}
C_{[\alpha\beta]\gamma\delta}=C_{\alpha\beta\gamma\delta}=
C_{\gamma\delta\alpha\beta}, \quad C^{\alpha}_{\phantom{\alpha}\alpha\gamma\delta}=0, \quad  C_{\alpha[\beta\gamma\delta]}=0.
\label{weyl_candidatep}
\end{equation}
%\label{weyl_candidate}
\noindent
From (\ref{weyl_candidatep}) we easily deduce that $\
^{*}C_{\alpha\beta\gamma\delta}=C^{*}_{\alpha\beta\gamma\delta}=
(1/2)\eta_{\gamma\delta\sigma\tau}
\dnup{C}{\alpha\beta}{\sigma\tau}$
 and indeed
$C^{*}_{\alpha\beta\gamma\delta}$ is also a Weyl candidate.
Given a Weyl candidate $C_{\alpha\beta\gamma\delta}$, its \emph{Weyl current}
$J_{\alpha\beta\gamma}$ is defined by
\begin{equation}
J_{\beta\gamma\delta}\equiv\nabla_{\alpha}C^{\alpha}_{\phantom{\alpha}\beta\gamma\delta}.
\label{weyl_current}
\end{equation}
The tensor $J_{\beta\gamma\delta}$ is trace-free, antisymmetric in
the last pair of indices and it has the property
$J_{[\beta\gamma\delta]}=0$. From (\ref{weyl_current}) we deduce
the following identities
\begin{equation}
 \nabla_{[\alpha}C_{\beta\gamma]\mu\nu}=
-\frac{1}{3}\eta_{\alpha\beta\gamma\sigma}J^{*\sigma}_{\phantom{*\sigma}\mu\nu},\quad
\nabla_{\beta}J^{\beta}_{\phantom{\beta}\gamma\delta}=
R_{[\gamma}^{\phantom{[\gamma}\sigma\rho\mu}C_{\delta]\sigma\rho\mu},
\label{weyl_bianchi}
\end{equation}
where $J^{*\sigma}_{\phantom{*\sigma}\mu\nu}\equiv
\eta_{\mu\nu}^{\phantom{\mu\nu}\rho\lambda}J^{\sigma}_{\phantom{\sigma}\rho\lambda}/2$.
Equations (\ref{weyl_current})-(\ref{weyl_bianchi}) have counterparts involving
$C^{*}_{\mu\nu\rho\sigma}$ which we will not discuss. Also, if we combine (\ref{weyl_current}) and (\ref{weyl_bianchi}) we get a wave equation for the Weyl candidate:
\begin{eqnarray}
\fl\nabla_{\alpha}\nabla^{\alpha}C_{\beta\gamma\chi\rho}=
-2R_{\gamma\phantom{\alpha}[\rho}^{\phantom{\gamma}\alpha\phantom{[\rho}\delta}
C_{\chi]\delta\beta\alpha}-2R_{\chi\phantom{\alpha}\rho}^{\phantom{\chi}\alpha
\phantom{\rho}\delta}C_{\beta\gamma\alpha\delta}+
2R_{[\rho}^{\phantom{[\rho}\alpha}C_{\chi]\alpha\beta\gamma}+
2R_{\beta\phantom{\alpha}[\rho}^{\phantom{\beta}\alpha\phantom{[\rho}\delta}
C_{\chi]\delta\gamma\alpha} \nonumber\\
\fl \hspace{3cm}+\eta_{\chi\rho\lambda\pi}\nabla^{\pi}
J^{*\lambda}_{\phantom{*\lambda}\beta\gamma}+2\nabla_{[\chi}J_{\rho]\beta\gamma}.
\label{box_weyl}
\end{eqnarray}
The previous considerations hold regardless of whether the Weyl
candidate is real or {\em complex}. Indeed, from any real Weyl
candidate $C_{\mu\nu\rho\sigma}$ we may construct a complex Weyl
candidate $\mathcal{C}_{\mu\nu\rho\sigma}$ by means of
\begin{equation}
\mathcal{C}_{\mu\nu\rho\sigma}=\frac{1}{2}(C_{\mu\nu\rho\sigma}+
\mbox{i}\ C^*_{\mu\nu\rho\sigma}).
\label{complexification}
\end{equation}
A Weyl candidate constructed in such a way is {\em
  self-dual}\footnote{Note that sometimes the opposite convention is
  followed in the literature whereby the complex conjugate of
  $\mathcal{C}_{\mu\nu\rho\sigma}$ is given the name of
  ``self-dual''.}
\begin{equation}
\mathcal{C}^*_{\mu\nu\rho\sigma}=\mbox{i}\
\mathcal{C}_{\mu\nu\rho\sigma}.
\label{self-dual}
\end{equation}
Reciprocally, if a complex Weyl candidate
$\mathcal{C}_{\mu\nu\rho\sigma}$ fulfils (\ref{self-dual}) then its
real part $\mbox{Re}(\mathcal{C}_{\mu\nu\rho\lambda})$ and its
imaginary part $\mbox{Im}(\mathcal{C}_{\mu\nu\rho\lambda})$ are also
Weyl candidates related via
\begin{equation}
\mbox{Im}(\mathcal{C}_{\mu\nu\rho\lambda})=
\frac{1}{2}\eta_{\rho\lambda\alpha\beta}\mbox{Re}(\mathcal
{C}_{\mu\nu}^{\phantom{\mu\nu}\alpha\beta}),\quad \mbox{Re}(\mathcal{C}_{\mu\nu\rho\lambda})=-
\frac{1}{2}\eta_{\rho\lambda\alpha\beta}\mbox{Im}(\mathcal
{C}_{\mu\nu}^{\phantom{\mu\nu}\alpha\beta}).
\label{reim}
\end{equation}

\subsection{The causal propagation of a Weyl candidate}
\label{causal-prop-weyl}
Besides the basic algebraic properties of a Weyl candidate just explained we need to introduce a further concept which will play a crucial role in the sequel.
\begin{definition}\label{causal-prop}
Let $(\mathcal{M},g_{\mu\nu})$ be a spacetime and consider a Weyl candidate
$C_{\alpha\beta\mu\nu}$ defined in the whole of $\mathcal{M}$.
We say that the Weyl candidate $C_{\alpha\beta\mu\nu}$ propagates causally
on $\mathcal{M}$ if for any embedded spacelike hypersurface $\mathcal{B}\subset\mathcal{M}$ the condition
$C_{\alpha\beta\mu\nu}|_{\mathcal{B}}=0$ implies $C_{\alpha\beta\mu\nu}=0$ on
$D(\mathcal{B})$.
\end{definition}

\medskip
\noindent
Given a set $\mathcal{B}$, a point $p\in\mathcal{M}$ belongs to the future Cauchy development $D^+(\mathcal{B})$ if any past-inextendible causal curve containing $p$ intersects $\mathcal{B}$. There is a similar notion of $D^-(\mathcal{B})$ and the set $D(\mathcal{B})\equiv D^+(\mathcal{B})\cup D^-(\mathcal{B})$ is called the {\em Cauchy development} of $\mathcal{B}$. It is well-known
that the interior of $D(\mathcal{B})$ is globally hyperbolic and since
$\mathcal{B}$ is an embedded spacelike hypersurface then it will be a Cauchy
hypersurface for $D(\mathcal{B})$ ---see e.g. \cite{WALD} for an
elementary introduction to these concepts of causality theory.

The simplest example of a Weyl candidate which propagates causally is
the Weyl tensor $W_{\alpha\beta\mu\nu}$ of a vacuum spacetime ---see
e.g. \cite{BONILLA}. It is possible to address the causal propagation
of a Weyl candidate by following the general techniques given in
\cite{BERGQVIST} ---see also \cite{HYPERBOLIZATION}--- in which the causal
propagation of any tensor is analysed. These ideas are reviewed next.

An essential object to study the causal propagation of a Weyl
candidate is its {\em Bel-Robinson tensor}.  Given a real Weyl
candidate $C_{\alpha\beta\gamma\delta}$, we define its Bel-Robinson
tensor by
\begin{equation}
T_{\alpha\beta\gamma\delta}\equiv
C_{\alpha\phantom{\rho}\delta}^{\phantom{\alpha}\rho\phantom{\delta}\mu}
C_{\beta\rho\gamma\mu}+
C_{\alpha\phantom{\sigma}\gamma}^{\phantom{\alpha}\mu\phantom{\gamma}\rho}
C_{\beta\mu\delta\rho}
-\frac{1}{8}g_{\alpha\beta}g_{\gamma\delta}C_{\mu\nu\rho\lambda}
C^{\mu\nu\rho\lambda}.
\label{b-r}
\end{equation}
The previous definition has been described in a more general framework
in \cite{SUPERENERGY} as the {\em superenergy tensor} of the Weyl
candidate $C_{\alpha\beta\gamma\delta}$. In our particular context we
prefer to call this tensor the Bel-Robinson tensor of
$C_{\alpha\beta\gamma\delta}$ in analogy with the Bel-Robinson tensor
constructed outof the Weyl tensor \cite{BEL}. The Bel-Robinson tensor
of a Weyl candidate has the following properties ---see
\cite{SUPERENERGY} for detailed proofs of points (\ref{first}), (\ref{second}) and (\ref{third}); point (\ref{fourth})
follows by taking the covariant divergence of (\ref{b-r}) and then using repeatedly (\ref{weyl_current}) and (\ref{weyl_bianchi}).
\begin{theorem}\label{sproperties}
If $T_{\alpha\beta\mu\nu}$ is the Bel-Robinson tensor of the Weyl candidate
$C_{\alpha\beta\mu\nu}$ then
\begin{enumerate}
 \item\label{first} $T_{(\alpha\beta\mu\nu)}=T_{\alpha\beta\mu\nu}$,
$T^{\alpha}_{\phantom{\alpha}\alpha\mu\nu}=0$.
\item\label{second}Generalized dominant property: if $u^{\mu}_1$, $u^{\mu}_2$, $u^{\mu}_3$, $u^{\mu}_4$ are causal future-directed vectors then $T_{\alpha\beta\mu\nu}u^{\alpha}_1u^{\beta}_2u^{\mu}_3u^{\nu}_4\geq 0$.
This property admits an alternative formulation: let $E=\{e_0^\mu,e_1^\mu,e_2^\mu,e_3^\mu\}$ be any orthonormal frame with $e_0^\mu$ the timelike vector. Then
the following inequality holds
$$
T_{0000}\geq |T_{\boldsymbol{a}\boldsymbol{b}\boldsymbol{c}\boldsymbol{d}}|,
$$
where the component indices refer to the frame $E$.
\item \label{third} $T_{\alpha\beta\mu\nu}=0\Longleftrightarrow C_{\alpha\beta\mu\nu}=0 \Longleftrightarrow\exists$ a timelike vector $u^{\mu}$ such that
$T_{\alpha\beta\mu\nu}u^{\alpha}u^{\beta}u^{\mu}u^{\nu}=0$.
\item \label{fourth}The covariant divergence of $T_{\alpha\beta\mu\nu}$ is given by
\begin{equation}
 \nabla_{\alpha}T^{\alpha}_{\phantom{\alpha}\beta\gamma\delta}=
4J^{\alpha\phantom{(\gamma}\sigma}_{\phantom{\alpha}(\gamma}
C_{\beta)\alpha\delta\sigma}-2J_{(\gamma}^{\phantom{(\gamma}\alpha\sigma}
C_{\beta)\delta\alpha\sigma}-g_{\beta\gamma}J^{\alpha\sigma\rho}
C_{\delta\alpha\sigma\rho}.
\label{covdivbr}
\end{equation}
\end{enumerate}
\end{theorem}

\begin{figure}[t]
\label{figure1}
\begin{center}
 \includegraphics[width=.5\textwidth,bb=0 0 549 401]{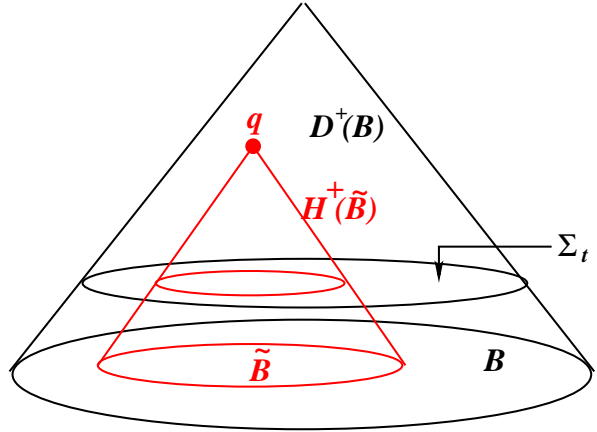}
 % CausalProp.eps: 0x0 pixel, 300dpi, 0.00x0.00 cm, bb=0 0 549 401
\end{center}
\caption{Schematic depiction of the sets $\mathcal{B}$,
  $\tilde{\mathcal{B}}$, $D^+(\mathcal{B})$ and
  $H^+(\tilde{\mathcal{B}})$ ---see the main text for further details.}
\end{figure}

Next, we explain how the properties summarized in theorem
\ref{sproperties} are useful in the study of the causal propagation of
$C_{\mu\nu\rho\lambda}$. We follow closely \cite{BERGQVIST} in our
exposition. Let $D^+(\mathcal{B})$ be the future Cauchy development of
a closed achronal hypersurface $\mathcal{B}$ and pick up
any event $q\in D^+(\mathcal{B})$. Furthermore, let $J^-(q)$ denote
the causal past of the event $q$. The set
$\mathcal{K}=J^-(q)\cap\overline{D^+(\mathcal{B})}$ is compact and it
contains the set
$\tilde{\mathcal{B}}\equiv\mathcal{K}\cap\mathcal{B}$.  The boundary
of $\mathcal{K}$ is $\partial\mathcal{K}=\tilde{\mathcal{B}}\cup
H^+(\tilde{\mathcal{B}})$ where $H^+(\tilde{\mathcal{B}})$ is the
future Cauchy horizon of $\tilde{\mathcal{B}}$. Also, since the
interior of $D^+(\mathcal{B})$ is globally hyperbolic we deduce that
it can be foliated by a family of spacelike hypersurfaces
$\{\Sigma_t\}$, $t\in I$, where $t_0\leq t\leq t_1$, $t_0<0$, $0<t_1$
and $\Sigma_0=\mathcal{B}$. The figure \ref{figure1} shows a schematic view
of this geometric construction.

Let $n_\mu$ be the
timelike normal 1-form to the leaves of $\{\Sigma_t\}_{t\in I}$ and
define the quantity
\begin{equation}
\fl W(t)\equiv\int_{J^{-}(\Sigma_{t})\cap\mathcal{K}}
T_{\mu\nu\alpha\beta}n^{\mu}n^{\nu}n^{\alpha}n^{\beta}\ d\mathcal{M}=
\int_{0}^tdt'\left(\int_{\Sigma_{t'}\cap\mathcal{K}}
T_{\mu\nu\alpha\beta}n^{\mu}n^{\nu}n^{\alpha}n^{\beta}\ d\Sigma_{t'}\right).
\end{equation}
Here, $d\mathcal{M}$ is the positive measure on the spacetime
$\mathcal{M}$ constructed from the volume form
$\eta_{\mu\nu\rho\sigma}$ (any measure defined from a volume form will be understood as positive) and $d\Sigma_{t'}$ is the measure on the
hypersurface $\Sigma_{t'}$ obtained from its volume form
$n^\mu\eta_{\mu\nu\rho\sigma}|_{\Sigma_{t'}}$.  The scalar
$T_{\mu\nu\rho\lambda}n^\mu n^\nu n^\rho n^\lambda$ is called
the \emph{superenergy density} and is everywhere non-negative ---see point
(\ref{second}) of theorem \ref{sproperties}--- which entails $W(t)\geq
0$.  Also, from the definition of $W(t)$ we get
\begin{equation}
W'(t)=\int_{\Sigma_{t}\cap\mathcal{K}}
T_{\mu\nu\alpha\beta}n^{\mu}n^{\nu}n^{\alpha}n^{\beta}\ d\Sigma_{t}=
\int_{\Sigma_{t}\cap\mathcal{K}}P_{\mu}n^{\mu}\ d\Sigma_{t},
\label{dwdt}
\end{equation}
where $P_\mu$ is defined by
\begin{equation}
P_{\mu}\equiv T_{\mu\nu\rho\sigma}n^{\nu}n^{\rho}n^{\sigma}.
\end{equation}
Again point (\ref{second}) of theorem \ref{sproperties} implies that
$P_\mu n^\mu$ is also a non-negative quantity and hence $W'(t)\geq 0$.
Now, according to the Gauss theorem we have
\begin{equation}
\fl\int_{J^-(\Sigma_{t})\cap\mathcal{K}}\nabla_{\mu}P^\mu\ d\mathcal{M}=
\int_{\Sigma_{t}\cap\mathcal{K}}P_{\mu}n^{\mu}\ d\Sigma_{t}-
\int_{\tilde{\mathcal{B}}}P_{\mu}n^{\mu}\ d\tilde{\mathcal{B}}+
\int_{H^+(\tilde{\mathcal{B}})}P_\mu k^\mu\ d(H^+(\tilde{\mathcal{B}})),
\label{gauss}
\end{equation}
where $d\tilde{\mathcal{B}}$ and $d(H^+(\tilde{\mathcal{B}}))$ denote
the measures on the hypersurfaces $\tilde{\mathcal{B}}$ and
$H^+(\tilde{\mathcal{B}})$ respectively induced by the volume forms
$n^\mu\eta_{\mu\nu\rho\lambda}|_{\tilde{\mathcal{B}}}$, and $k^\mu\eta_{\mu\nu\rho\lambda}|_{H^+(\tilde{\mathcal{B}})}$. In this last expression $k^\mu$ is a causal vector which is orthogonal to the null hypersurface $H^+(\tilde{\mathcal{B}})$, points outward to $\mathcal{K}$ (therefore it is future-directed) and is such that $k^\mu\eta_{\mu\nu\rho\lambda}|_{H^+(\tilde{\mathcal{B}})}\neq 0$.
Note that these three conditions do not fix the vector $k^\mu$ univocally.
By combining (\ref{dwdt}) and (\ref{gauss}) we deduce
\begin{equation}
\fl W'(t)=\int_{J^-(\Sigma_{t})\cap\mathcal{K}}\nabla_{\mu}P^\mu\ d\mathcal{M}+
\int_{\tilde{\mathcal{B}}}P^{\mu}n_{\mu}\ d\tilde{\mathcal{B}}-
\int_{H^+(\tilde{\mathcal{B}})\cap J^-(\Sigma_t)}P^\mu k_\mu\ d(H^+(\tilde{\mathcal{B}})).
\label{estimate1}
\end{equation}
Now, we need to estimate the right hand side of (\ref{estimate1}).
According to point (\ref{second}) of theorem \ref{sproperties}, the quantity
$P^\mu k_\mu$ is positive and therefore
\begin{equation}
0\leq\int_{H^+(\tilde{\mathcal{B}})\cap J^-(\Sigma_t)}P^\mu k_\mu\ d(H^+(\tilde{\mathcal{B}})).
\end{equation}
On the other hand, if $C_{\alpha\beta\mu\nu}$ vanishes on $\mathcal{B}$ then
$W(0)=0$ and
\begin{equation}
\int_{\tilde{\mathcal{B}}}P^{\mu}n_{\mu}\ d\tilde{\mathcal{B}}=0.
\end{equation}
Next, we set up an orthonormal frame
$E=\{e_0^\mu,e_1^\mu,e_2^\mu,e_3^\mu\}$ with $e_0^\mu=n^\mu$ and we
consider the set of components
$\nabla_{\boldsymbol{a}}n_{\boldsymbol{b}}$ in such a frame. Since
these are continuous functions we deduce that in the compact set
$\mathcal{K}$ the estimate
$\nabla_{\boldsymbol{a}}n_{\boldsymbol{b}}\leq m_0$,
$\forall\boldsymbol{a},\boldsymbol{b}=0,1,2,3$ holds for some constant
$m_0$. Hence
\begin{eqnarray}
\fl\nabla_{\boldsymbol{a}}P^{\boldsymbol{a}}=3n_{\boldsymbol{c}}n_{\boldsymbol{d}}T^{\boldsymbol{cdab}}\nabla_{\boldsymbol{a}}
n_{\boldsymbol{b}}+
n_{\boldsymbol{a}}
n_{\boldsymbol{b}}n_{\boldsymbol{c}}\nabla_{\boldsymbol{d}}T^{\boldsymbol{d abc}}\nonumber\\
\fl \hspace{1cm}\leq 3m_0
\sum^{3}_{\boldsymbol{c},\boldsymbol{d}=0}T^{\boldsymbol{00cd}}+
n_{\boldsymbol{a}}
n_{\boldsymbol{b}}n_{\boldsymbol{c}}\nabla_{\boldsymbol{d}}T^{\boldsymbol{d
abc}}\leq 3m_0
T^{\boldsymbol{abcd}}n_{\boldsymbol{a}}n_{\boldsymbol{b}}
n_{\boldsymbol{c}}n_{\boldsymbol{d}}+ n_{\boldsymbol{a}}
n_{\boldsymbol{b}}n_{\boldsymbol{c}}\nabla_{\boldsymbol{d}}T^{\boldsymbol{dabc}}, \nonumber \\
&&
\end{eqnarray}
where in the last step, point (\ref{second}) of theorem \ref{sproperties} was used. From here, we deduce
\begin{eqnarray}
\int_{J^-(\Sigma_{t})\cap\mathcal{K}}\nabla_{\mu}P^\mu\ d\mathcal{M} \nonumber \\
\fl\leq \int_{J^-(\Sigma_{t})\cap\mathcal{K}}(\nabla_{\mu}T^{\mu\nu\rho\sigma})n_{\nu}n_{\rho}n_{\sigma}\ d\mathcal{M}+3m_0
\int_{J^-(\Sigma_{t})\cap\mathcal{K}}T^{\mu\nu\rho\sigma}n_{\mu}n_{\nu}n_{\rho}n_{\sigma}\ d\mathcal{M}.
\end{eqnarray}
Using the above estimates in (\ref{estimate1}) we get
\begin{equation}
0\leq W'(t)\leq \int_{J^-(\Sigma_{t})\cap\mathcal{K}}
(\nabla_{\mu}T^{\mu\nu\rho\sigma})n_{\nu}n_{\rho}n_{\sigma}\ d\mathcal{M}+3m_0 W(t).
\label{gronwall}
\end{equation}
Now it only remains to estimate
$n_{\nu}n_{\rho}n_{\sigma}\nabla_{\mu}T^{\mu\nu\rho\sigma}$ and to
that end (\ref{covdivbr}) is used. The actual estimate will depend on
the particular expression of the Weyl current. For example, the
simplest case arises when $J_{\mu\nu\rho}=0$ where no further
estimates are needed and Gr\"onwall's lemma entails
\begin{equation}
W(t)=0,\quad  t\in [0,t_1),
\end{equation}
which in turn implies that the scalar
$T_{\mu\nu\alpha\beta}n^{\mu}n^{\nu}n^{\alpha}n^{\beta}$ vanishes in
$\mathcal{K}$. From the arbitrariness of $q$ we deduce that the
superenergy density is zero in $D^+(\mathcal{B})$ and by a similar
argument as the above one can show that this is also the case for $D^-(\mathcal{B})$.
Hence point 3 of theorem \ref{sproperties} implies that
$C_{\alpha\beta\mu\nu}=0$ on $D^\pm(\mathcal{B})$. Consequently the tensor
$C_{\alpha\beta\mu\nu}$ propagates causally whenever the Weyl current is
zero.
\begin{remark}\em
  Point (\ref{third}) of theorem \ref{sproperties} entails
  $C_{\alpha\beta\mu\nu}|_{\mathcal{B}}=0$ if and only if
  $(T_{\alpha\beta\mu\nu}n^{\alpha}n^{\beta}n^{\mu}n^{\nu})|_{\mathcal{B}}=0$.
  Accordingly, the causal propagation of a Weyl candidate can also be
  formulated in terms of the vanishing of the superenergy density at
  the hypersurface $\mathcal{B}$.
\label{remark:se_density}
\end{remark}

\section{The Mars-Simon tensor}\label{section:mstensor}
We start this section by recalling some basic background about vacuum
spacetimes possessing a Killing vector ---see e.g. \cite{HEUSLER} for
a complete account. Let $(\mathcal{M},g_{\mu\nu})$ be a vacuum
solution of Einstein field equations and let $\xi^\mu$ be a Killing
vector, $\nabla_\mu\xi_\nu + \nabla_\nu \xi_\mu=0$. This condition
enables us to define a 2-form, the \emph{Killing form}, by
\begin{equation}
F_{\mu\nu}\equiv\nabla_{[\mu}\xi_{\nu]} = \nabla_\mu\xi_\nu.
\label{killing-2form}
\end{equation}
Elementary manipulations lead to
\begin{equation}
\nabla_{\mu}F_{\nu\lambda}=W_{\nu\lambda\mu\rho}\xi^\rho.
\label{covdivF}
\end{equation}
The \emph{self-dual Killing form} $\mathcal{F}_{\mu\nu}$ is then
\begin{equation}
\mathcal{F}_{\mu\nu} =F_{\mu\nu}+\mbox{i} F^*_{\mu\nu},
\label{mcalf}
\end{equation}
with $F^*_{\mu\nu}=(1/2) \eta_{\mu\nu\lambda\rho}F^{\lambda\rho}$ the
Hodge dual of $F_{\mu\nu}$. From the previous definition it is
straightforward to obtain the property (self-duality)
\begin{equation}
\mathcal{F}_{\mu\nu}=\mbox{i}\ \mathcal{F}^*_{\mu\nu}.
\label{self-duality}
\end{equation}
We shall write $\mathcal{F}^2= \mathcal{F}_{\mu\nu}
\mathcal{F}^{\mu\nu}$. The Killing form is said to be degenerate if
$\mathcal{F}^2=0$. Other basic algebraic properties of
$\mathcal{F}_{\mu\nu}$ are
\begin{equation}
\mathcal{F}_{\mu\rho}\mathcal{F}_{\nu}^{\phantom{\nu}\rho}=
\frac{1}{4}g_{\mu\nu}\mathcal{F}^2,\quad \mathcal{F}_{[\mu}^{\phantom{\mu}\rho}\overline{\mathcal{F}}_{\nu]\rho}=0,\quad
\mathcal{F}_{\mu\nu}\overline{\mathcal{F}}^{\mu\nu}=0.
\label{f2}
\end{equation}
From a Killing vector $\xi^\mu$ we may define its associated
\emph{Ernst 1-form} by
\begin{equation}
\sigma_\nu\equiv 2 \xi^\mu \mathcal{F}_{\mu\nu}= -\nabla_\nu \Lambda - \mbox{i} \Omega_\nu,
\label{ernst1}
\end{equation}
where $\Lambda\equiv\xi_\mu \xi^\mu$ is the norm of the Killing vector
and $\Omega_\mu\equiv\eta_{\mu\nu\lambda\rho} \xi^\nu \nabla^\lambda
\xi^\rho$ denotes its \emph{twist form}. It is well-known that in a
vacuum spacetime $\sigma_\mu$ is closed ---i.e.
$\nabla_{[\nu}\sigma_{\mu]}=0$. Thus locally ---if we are in a simply
connected region--- there exists a scalar field $\sigma$, the
\emph{Ernst potential}, such that $\sigma_\mu =\nabla_\mu \sigma$. The
self-dual Weyl tensor and the self-dual Killing form are related by
\begin{equation}
\nabla_{\mu}\mathcal{F}_{\nu\rho}=2\overline{\mathcal{W}}_{\nu\rho\mu\alpha}
\xi^\alpha,
\label{covdivFS}
\end{equation}
a property which is easily obtained from (\ref{covdivF}). Finally, we
introduce the tensor $\mathcal{I}_{\mu\nu\lambda\rho}$ defined by
\begin{equation}
\mathcal{I}_{\mu\nu\lambda\rho}\equiv\frac{1}{2} \left(g_{\mu\lambda}g_{\nu\rho}-g_{\mu\rho}g_{\nu\lambda}+\mbox{i} \eta_{\mu\nu\lambda\rho}\right).
\label{2formelement}
\end{equation}
The tensor $\mathcal{I}_{\mu\nu\lambda\rho}$ has the
obvious symmetries $\mathcal{I}_{\mu\nu\lambda\rho}=\mathcal{I}_{[\mu\nu]\lambda\rho}=
\mathcal{I}_{\lambda\rho\mu\nu}$. This tensor
can be regarded as a metric in the space of self-dual 2-forms.

We have now all the ingredients to define the \emph{Mars-Simon
  tensor}. This is a four-rank tensor
$\mathcal{S}_{\mu\nu\lambda\rho}$ whose expression is given by
\footnote{We follow the convention set in \cite{IONESCU} of calling
  this tensor the Mars-Simon tensor. Note however, that our
  conventions are slightly different.}
\begin{equation}
\mathcal{S}_{\mu\nu\lambda\rho}\equiv 2\overline{\mathcal{W}}_{\mu\nu\lambda\rho} + \frac{6}{1-\sigma} \left( \mathcal{F}_{\mu\nu}\mathcal{F}_{\lambda\rho} -\frac{1}{6} \mathcal{F}^2 \mathcal{I}_{\mu\nu\lambda\rho}   \right). \label{Mars-Simon:tensor}
\end{equation}
From the previous expression we deduce that in a vacuum spacetime
admitting a Killing vector, the Mars-Simon tensor is determined up
to a choice for the Ernst potential. We make a choice such that
$\sigma\neq 1$ on $\mathcal{M}$. Our interest in the Mars-Simon
tensor lies in the following result ---see \cite{MARS-1,MARS-2}
for a proof.
\begin{theorem}
Let $(\mathcal{M},g_{\mu\nu})$ be a smooth, vacuum spacetime with a Killing vector $\xi^\mu$. Let $\mathcal{F}_{\mu\nu}$ denote the associated self-dual Killing form. If there is a non-vanishing real constant $M$  such that the conditions
\numparts
\begin{eqnarray}
&& \mathcal{F}^2=-M^2(1-\sigma)^4,\label{kerr_condition1} \\
&& \mathcal{S}_{\mu\nu\lambda\rho}=0\label{kerr_condition2},
\end{eqnarray}
\endnumparts
hold on a non-empty $\mathcal{N}\subset\mathcal{M}$
then $(\mathcal{N},g_{\mu\nu})$ is locally isometric to the Kerr spacetime.
\label{kerr-characterization}
\end{theorem}

Condition (\ref{kerr_condition1}) of theorem
\ref{kerr-characterization} can be reformulated in the form
\begin{equation}
\Xi(M,\vec{\xi})=0,
\end{equation}
where the scalar $\Xi(M,\vec{\xi})$ is defined for any Killing vector $\vec{\xi}$ and any real constant $M$ by the expression
\begin{equation}
\Xi(M,\vec{\xi})\equiv\mathcal{F}^2+M^2(1-\sigma)^4. \label{Mars-Simon:modified}
\end{equation}
A very important property of the scalar $\Xi(M,\vec{\xi})$ is that it fulfils the differential equation
\begin{equation}
 \nabla_\mu\Xi=-2\mathcal{F}^{\alpha\beta}S_{\mu\rho\alpha\beta}\xi^\rho+
\frac{4\Xi\sigma_\mu}{\sigma-1}.
\label{CDXi}
\end{equation}
To see it, differentiate both sides of (\ref{Mars-Simon:modified}) and use (\ref{covdivFS}) to replace the covariant derivatives of $\mathcal{F}_{\mu\nu}$. The result is
\begin{equation}
 \nabla_\mu\Xi=-4(\mathcal{F}^{\alpha\beta}\overline{\mathcal{W}}_{\mu\rho\alpha\beta}\xi^\rho+M^2\sigma_\mu(1-\sigma)^3).
\end{equation}
Next use (\ref{Mars-Simon:tensor}) to write $\overline{\mathcal{W}}_{\mu\rho\alpha\beta}$ in terms of $\mathcal{S}_{\mu\rho\alpha\beta}$ and replace all the occurrences of
$\mathcal{I}_{\mu\rho\alpha\beta}$ by means of (\ref{2formelement}) to obtain
\begin{equation}
 \nabla_\mu\Xi=2(2M^2\sigma_\mu(\sigma-1)^3-\mathcal{F}^{\alpha\beta}
S_{\mu\rho\alpha\beta}\xi^\rho)+\frac{\mathcal{F}^2(-10\mathcal{F}_{\mu\rho}+2\mbox{i}\ \mathcal{F}^*_{\mu\rho})\xi^\rho}{\sigma-1}.
\end{equation}
We use here (\ref{self-duality}) and (\ref{ernst1}) with the result
\begin{equation}
\nabla_\mu\Xi=-2\mathcal{F}^{\alpha\beta}S_{\mu\rho\alpha\beta}\xi^\rho+\frac{4\sigma_\mu(\mathcal{F}^2+M^2(1-\sigma)^4)}{\sigma-1},
\end{equation}
from which (\ref{CDXi}) is deduced.

\subsection{Mars-Simon tensor as a Weyl candidate}
In \cite{IONESCU} it is proved that the Mars-Simon tensor
$\mathcal{S}_{\mu\nu\alpha\beta}$ is a Weyl candidate. Furthermore, it
was shown that its Weyl current is a linear function of
$\mathcal{S}_{\mu\nu\alpha\beta}$ in which case the wave equation
(\ref{box_weyl}) is homogeneous in $\mathcal{S}_{\mu\nu\alpha\beta}$.
The derivation of these results are discussed as some of the
intermediate calculations will be needed in section \ref{cp-ms}.
\begin{proposition}
The Mars-Simon tensor is a Weyl candidate.
\label{mars-weyl}
\end{proposition}
\noindent
{\it Proof.\hspace{.2cm}} From (\ref{Mars-Simon:tensor}) it
is clear that
$\mathcal{S}_{[\mu\nu]\alpha\lambda}=\mathcal{S}_{\mu\nu[\alpha\lambda]}=
\mathcal{S}_{\alpha\lambda\mu\nu}$. Using (\ref{f2}) one can show that
$\mathcal{S}_{\mu\rho\phantom{\rho}\lambda}^{\phantom{\mu\rho}\rho}=0$
and therefore the Mars-Simon tensor is traceless. To finish our proof
we need to show that the Mars-Simon tensor fulfils the cyclic
identity. To that end we compute $\mathcal{S}_{\mu[\nu\rho\lambda]}$
getting
\begin{equation}
\mathcal{S}_{\mu[\nu\rho\lambda]}=\frac{6}{1-\sigma}
\mathcal{F}_{\mu[\nu}\mathcal{F}_{\rho\lambda]}-\frac{\mbox{i}\ \mathcal{F}^2}{2(1-\sigma)}
\eta_{\mu\nu\rho\lambda}.
\end{equation}
The right hand side of this expression vanishes by virtue of the identity
\begin{equation}
\mathcal{F}_{\mu[\nu}\mathcal{F}_{\rho\lambda]}=\frac{1}{12}\mbox{i}\ \mathcal{F}^2
\eta_{\mu\nu\rho\lambda},
\end{equation}
which in turn is a consequence of the algebraic property $\mathcal{F}_{\mu[\nu}\mathcal{F}_{\rho\lambda]}=
\mathcal{F}_{[\mu\nu}\mathcal{F}_{\rho\lambda]}$ and the fact that in four dimensions
$\mathcal{F}_{[\mu\nu}\mathcal{F}_{\rho\lambda]}$ is proportional to $\eta_{\mu\nu\rho\lambda}$.\qed

\bigskip
A direct computation shows that the Mars-Simon tensor is self-dual
\begin{equation}
\mathcal{S}_{\mu\nu\rho\lambda}=\mbox{i}\ \mathcal{S}^*_{\mu\nu\rho\lambda}.
\end{equation}
Therefore $\mbox{Re}(\mathcal{S}_{\mu\nu\rho\lambda})$ and
$\mbox{Im}(\mathcal{S}_{\mu\nu\rho\lambda})$ are then related by duality
as explained in section \ref{weyl:c}:
\begin{equation}
\mbox{Im}(\mathcal{S}_{\mu\nu\rho\lambda})=
\frac{1}{2}\eta_{\rho\lambda\alpha\beta}\mbox{Re}(\mathcal
{S}_{\mu\nu}^{\phantom{\mu\nu}\alpha\beta}),\quad \mbox{Re}(\mathcal{S}_{\mu\nu\rho\lambda})=-
\frac{1}{2}\eta_{\rho\lambda\alpha\beta}\mbox{Im}(\mathcal
{S}_{\mu\nu}^{\phantom{\mu\nu}\alpha\beta}).
\label{re-im}
\end{equation}
Any property of $\mathcal{S}_{\mu\nu\rho\lambda}$ will admit an equivalent counterpart formulated in terms of either its real or imaginary part. In particular, this shows that both $\mbox{Re}(\mathcal{S}_{\mu\nu\rho\lambda})$ and $\mbox{Im}(\mathcal{S}_{\mu\nu\rho\lambda})$ are Weyl candidates and any of them contains the same information as the original Mars-Simon tensor.
\begin{proposition}\label{covdivms}
  The covariant divergence of the Mars-Simon tensor is given by
\begin{equation}
\nabla_{\alpha}\mathcal{S}^{\alpha}_{\phantom{\alpha}\beta\gamma\delta}=
\frac{1}{\sigma-1}\left(4(\mathcal{F}_{[\delta}^{\phantom{[\delta}\rho}
\mathcal{S}_{\gamma]\beta\alpha\rho}+\mathcal{F}_{\beta}^{\phantom{\beta}\rho}
\mathcal{S}_{\gamma\delta\alpha\rho})+2g_{\beta[\gamma}\mathcal{S}_{\delta]\alpha\lambda\rho}\mathcal{F}^{\lambda\rho}\right)\xi^\alpha.
\label{covdivmarssimon}
\end{equation}
\label{cdms}
\end{proposition}
\noindent
{\it Proof.\hspace{.2cm}}
To calculate
$\nabla_{\alpha}\mathcal{S}^{\alpha}_{\phantom{\alpha}\beta\gamma\delta}$
we start from (\ref{Mars-Simon:tensor}) and take the covariant
divergence of $\mathcal{S}_{\mu\nu\lambda\rho}$. Next we apply the
relations (\ref{covdivFS}), $\nabla_\mu\sigma=\sigma_\mu$,
$\nabla_\alpha\overline{\mathcal{W}}^\alpha_{\phantom{\alpha}\sigma\mu\nu}=0$
and $\nabla_{\alpha}\mathcal{I}_{\mu\nu\rho\sigma}=0$.  Finally we
expand all the occurrences of $\mathcal{I}_{\mu\nu\rho\sigma}$ using
(\ref{2formelement}) with the result
\begin{eqnarray}
&&\nabla_\alpha\mathcal{S}^\alpha_{\phantom{\alpha}\beta\gamma\delta}=\frac{1}{2(1-\sigma)^2}(-12\mathcal{F}_{\beta\alpha}\mathcal{F}_{\gamma\delta}\sigma^\alpha+
(\mbox{i}\ \eta_{\beta\gamma\delta\alpha}\sigma^\alpha+2g_{\beta[\gamma}\sigma_{\delta]})
\mathcal{F}^2) \nonumber \\
&&+\frac{1}{1-\sigma}((2\mbox{i}\ \eta_{\beta\gamma\delta\alpha}
\overline{\mathcal{W}}_{\rho\phantom{\alpha}\pi\kappa}^{\phantom{\rho}\alpha}
\xi^\rho+4g_{\beta[\delta}\overline{\mathcal{W}}_{\gamma]\alpha\pi\kappa}\xi^\alpha)\mathcal{F}^{\pi\kappa}-12\mathcal{F}_{\beta}^{\phantom{\beta}\rho}\overline{\mathcal{W}}_{\gamma\delta\alpha\rho}\xi^\alpha)
\end{eqnarray}
The dependence on the Ernst 1-form can be removed from this expression
if we write $\sigma_\mu$ in terms of $\mathcal{F}_{\mu\nu}$ and
$\xi^\mu$ by means of (\ref{ernst1}). The final step is to rearrange
the terms in the resulting expression in order to render it in the
form of (\ref{covdivmarssimon}). This is accomplished by using
(\ref{Mars-Simon:tensor}) to write
$\overline{\mathcal{W}}_{\mu\nu\lambda\rho}$ in terms of the
Mars-Simon tensor, expanding again all the occurrences of
$\mathcal{I}_{\mu\nu\rho\sigma}$ by means of (\ref{2formelement}) and
using (\ref{f2}), (\ref{self-duality}) where necessary.\qed

Equation (\ref{covdivmarssimon}) enables us to define the Weyl current
of the Mars-Simon tensor as
\begin{equation}
\mathcal{J}_{\beta\gamma\delta}\equiv\frac{1}{\sigma-1}\left(4(\mathcal{F}_{[\delta}^{\phantom{[\delta}\rho}
\mathcal{S}_{\gamma]\beta\alpha\rho}+\mathcal{F}_{\beta}^{\phantom{\beta}\rho}
\mathcal{S}_{\gamma\delta\alpha\rho})+2g_{\beta[\gamma}\mathcal{S}_{\delta]\alpha\sigma\rho}\mathcal{F}^{\sigma\rho}\right)\xi^\alpha.
\label{ms-weylcurrent}
\end{equation}

Since $\mathcal{J}_{\beta\gamma\delta}$ is linear in the Mars-Simon
tensor, we conclude ---in view of (\ref{box_weyl})--- that
$\mathcal{S}_{\mu\nu\rho\sigma}$ fulfils a homogeneous wave equation.
This interesting property was used in \cite{IONESCU} to show that the
{\em domain of outer communication} of an asymptotically flat smooth
stationary spacetime is, under certain additional conditions,
isometric to the Kerr domain of outer communication ---i.e. the
\emph{uniqueness of Kerr}.

Using $\nabla_\alpha\mathcal{S}^\alpha_{\phantom{\alpha}\mu\nu\rho}=
\mathcal{J}_{\mu\nu\rho}$ together with the self-duality of the
Mars-Simon tensor one deduces
\begin{equation}
\mathcal{J}_{\mu\nu\rho}=\mbox{i}\ \mathcal{J}^*_{\mu\nu\rho}=
\frac{\mbox{i}}{2}\eta_{\nu\rho\alpha\beta}\mathcal{J}_\mu^{\phantom{\mu}\alpha\beta},
\end{equation}
that is to say, $\mathcal{J}_{\mu\nu\rho}$ is also self-dual with
respect to the block of indices $\nu\rho$.

\subsection{Orthogonal splitting of the Mars-Simon tensor}
In our forthcoming calculations it is necessary to use the orthogonal
splitting of $\mathcal{S}_{\mu\nu\rho\lambda}$ with respect to a unit
timelike vector field $n^\mu$. Since $\mathcal{S}_{\mu\nu\rho\lambda}$
is a Weyl candidate we can use (\ref{decompose-weyl}) to find such a
orthogonal splitting. The resulting split is given by
\begin{equation}
\mathcal{S}_{\mu\nu\rho\lambda}=2(l_{\nu[\lambda}\mathcal{T}_{\rho]\mu}+
l_{\mu[\rho}\mathcal{T}_{\lambda]\nu}+\mbox{i}\ \varepsilon_{\rho\lambda\alpha}n_{[\mu}\mathcal{T}_{\nu]}^{\phantom{\nu]}\alpha}+
\mbox{i}\ \varepsilon_{\mu\nu\alpha}n_{[\rho}\mathcal{T}_{\lambda]}^{\phantom{\nu]}\alpha}),
\label{decompose-ms}
\end{equation}
where
\begin{equation}
\mathcal{T}_{\mu\nu}\equiv\mathcal{S}_{\mu\rho\nu\lambda}n^\rho n^\lambda.
\label{tspatial}
\end{equation}
The tensor $\mathcal{T}_{\mu\nu}$ is symmetric, spatial and traceless.
Alternatively, the orthogonal splitting of the Mars-Simon tensor can
be calculated directly from (\ref{Mars-Simon:tensor}). To that end one
needs to find the orthogonal splitting of the different quantities
which appear in (\ref{Mars-Simon:tensor}). The orthogonal splitting of
$\mathcal{W}_{\mu\nu\rho\lambda}$ is given in (\ref{decompose-weyl})
and the orthogonal splitting of $\mathcal{F}_{\mu\nu}$ is calculated
by means of (\ref{decompose_H}) and is given by
\begin{equation}
\mathcal{F}_{\mu\nu}=\mbox{i}\ \varepsilon_{\rho\mu\nu}\mathcal{E}^\rho-2\mathcal{E}_{[\mu}n_{\nu]},
\label{decompose-f}
\end{equation}
where
\begin{equation}
\mathcal{E}_\mu\equiv\mathcal{F}_{\mu\rho}n^\rho.
\label{define-e}
\end{equation}
One also needs the orthogonal splitting of
$\mathcal{I}_{\mu\nu\rho\lambda}$ which is easily calculated from
(\ref{2formelement}) and (\ref{decompose_eta}) so that
\begin{equation}
\mathcal{I}_{\mu\nu\rho\lambda}=h_{\mu[\rho}h_{\lambda]\nu}+2n_{[\nu}h_{\mu][\lambda}n_{\rho]}+\mbox{i}\ (n_{[\mu}\varepsilon_{\nu]\rho\lambda}+n_{[\rho}\varepsilon_{\lambda]\mu\nu}).
\label{decompose_I}
\end{equation}
We insert relations (\ref{decompose-weyl}), (\ref{decompose-f}) and
(\ref{decompose_I}) into the formula (\ref{Mars-Simon:tensor}) and
then equal the resulting expression to (\ref{decompose-ms}). The
result of that is the relation
\begin{equation}
\mathcal{T}_{\mu\nu}=2\overline{\mathcal{E}}_{\mu\nu}+\frac{2}{1-\sigma}(3\mathcal{E}_{\mu}
\mathcal{E}_{\nu}-h_{\mu\nu}\mathcal{E}_{\alpha}\mathcal{E}^{\alpha}),
\label{tgoestoe}
\end{equation}
which shall be used later on. Also needed later on is the relation
\begin{equation}
\mathcal{F}^2= -4\mathcal{E}_\mu\mathcal{E}^\mu
\Longrightarrow\Xi(M,\vec{\xi})=-4\mathcal{E}_\mu\mathcal{E}^\mu+M^2(1-\sigma)^4,
\label{ot-f2}
\end{equation}
which is derived from equation (\ref{decompose-f}).

The Killing vector $\xi^\mu$ can be decomposed in the form $\xi^\mu=-Yn^\mu+Y^\mu$ where $Y$, $Y^\mu$ are called the \emph{Killing lapse} and the \emph{Killing shift} respectively.
We need to find a formula relating $\mathcal{E}_\mu$ to $Y$, $Y_\mu$. To that end, we start by taking the covariant derivative of both sides of $\xi_\mu=-Yn_\mu+Y_\mu$ and replace $\nabla_\mu n_\nu$ by means of
(\ref{gradnormaltoextrinsic}). The result is
$$
\nabla_\mu\xi_\nu=Y(K_{\mu\nu}+A_{\nu}n_\mu)+\nabla_\mu Y_\nu-n_\nu\nabla_\mu Y.
$$
Next we use in this expression the property $\nabla_\mu Y=D_\mu Y-n_\mu\mathcal{L}_{\vec{n}}Y$ and equation (\ref{decompose_cdT}) to replace
$\nabla_\mu Y$ and $\nabla_\mu Y_\nu$ respectively yielding
\begin{eqnarray}
&&\nabla_\mu\xi_\nu=Y K_{\mu\nu}+D_\mu Y_\nu+n_\mu n_\nu(\mathcal{L}_{\vec{n}}Y-A^\rho Y_\rho)\nonumber\\
&&\hspace{2cm}+n_\mu(Y A_\nu-K_{\nu\rho}Y^\rho-\mathcal{L}_{\vec{n}}Y_\nu)-n_\nu(K_{\mu\rho}Y^\rho+D_\mu Y).\label{decompose_cdxi}
\end{eqnarray}
The antisymmetric part of this expression gives the
 orthogonal splitting of the 2-form $F_{\mu\nu}$ which is
\begin{equation}
F_{\mu\nu}=-Y A_{[\mu}n_{\nu]}+n_{[\mu}D_{\nu]}Y+D_{[\mu}Y_{\nu]}-
n_{[\mu}\mathcal{L}_{\vec{n}}Y_{\nu]}.
\label{ot-f}
\end{equation}
Also the symmetric part of (\ref{decompose_cdxi}) is the Killing condition $\nabla_{(\mu}\xi_{\nu)}=0$. Explicitly one has
\begin{eqnarray*}
\fl Y K_{\mu\nu}+D_{(\mu}Y_{\nu)}+n_\mu n_\nu(\mathcal{L}_{\vec{n}}Y-A^\rho Y_\rho)+
n_{(\mu}(A_{\nu)}Y-2K_{\nu)\rho}Y^\rho-D_{\nu)}Y-\mathcal{L}_{\vec{n}}Y_{\nu)})=0,
\end{eqnarray*}
from which we deduce
\begin{equation}
\mathcal{L}_{\vec{n}}Y_\mu=A_\mu Y-D_\mu Y-2K_{\mu\rho}Y^\rho,
\end{equation}
which renders the relation (\ref{ot-f}) in the form
\begin{equation}
 F_{\mu\nu}=2n_{[\mu}K_{\nu]\rho}Y^\rho+D_{[\mu}Y_{\nu]}+2n_{[\mu}D_{\nu]}Y.
\label{decompose_f}
\end{equation}
From this expression we may compute the orthogonal splitting of
$F^*_{\mu\nu}$ which is
\begin{equation}
F^*_{\mu\nu}=\varepsilon_{\mu\nu\beta}(K_{\alpha}^{\phantom{\alpha}\beta}Y^\alpha+D^\beta Y)+n_{[\mu}\varepsilon_{\nu]\alpha\beta}D^\alpha Y^\beta,
\label{decompose_fdual}
\end{equation}
where equation (\ref{decompose_eta}) was used along the way.
Inserting equations (\ref{decompose_f}) and (\ref{decompose_fdual}) in
(\ref{mcalf}) and applying (\ref{define-e}) on the resulting
expression we deduce
\begin{equation}
 \mathcal{E}_\mu=K_{\mu\rho}Y^\rho+D_\mu Y-\frac{1}{2}\mbox{i}\
\varepsilon_{\mu\nu\rho}D^{\nu}Y^{\rho},
\label{def-mce}
\end{equation}
which is the required relation.

\section{The causal propagation of the Mars-Simon tensor}
\label{cp-ms}
In this section a proof of the following result is provided.
\begin{theorem}
The Mars-Simon tensor $\mathcal{S}_{\mu\nu\rho\lambda}$ propagates causally.
\label{ms-propagation}
\end{theorem}

\noindent {\it Proof.} According to (\ref{re-im}) it is enough to show
that either $\mbox{Re}(\mathcal{S}_{\mu\nu\rho\lambda})$ or
$\mbox{Im}(\mathcal{S}_{\mu\nu\rho\lambda})$ propagates causally. We
choose to work with the former, so let us set
$S_{\mu\nu\rho\lambda}\equiv
\mbox{Re}(\mathcal{S}_{\mu\nu\rho\lambda})$ and define
$L_{\mu\nu\lambda}\equiv \mbox{Re}(\mathcal{J}_{\mu\nu\lambda})$.
Equation (\ref{covdivmarssimon}) entails
\begin{equation}
\nabla_{\alpha}S^{\alpha}_{\phantom{\alpha}\beta\gamma\delta}=
L_{\beta\gamma\delta}.
\end{equation}
Hence, $L_{\beta\gamma\delta}$ is the Weyl current of the Weyl
candidate $S_{\mu\nu\rho\lambda}$. To study the causal propagation of
$S_{\mu\nu\rho\lambda}$ we follow the general procedure explained in
section \ref{causal-prop-weyl}. We denote by $B_{\mu\nu\rho\lambda}$
the Bel-Robinson tensor constructed from the Weyl candidate
$S_{\mu\nu\rho\lambda}$. To prove our result, we need to find a good
estimate for the quantity $n^\nu n^\rho n^\lambda\nabla_\mu
B^\mu_{\phantom{\mu}\nu\rho\lambda}$ ---see equation
($\ref{gronwall}$). One can write
\begin{equation}
n^\nu n^\rho n^\lambda\nabla_\mu B^\mu_{\phantom{\mu}\nu\rho\lambda}=L^{\rho\nu\lambda}n^\mu S_{\mu\rho\nu\lambda}+4L^{\mu\phantom{\rho}\nu}_{\phantom{\mu}\rho}n^\rho n^\lambda n^\alpha S_{\lambda\mu\alpha\nu},
\label{q-est}
\end{equation}
where (\ref{covdivbr}) was used. In order to work out the right hand
side of equation (\ref{q-est}) we calculate the orthogonal splitting
of $S_{\mu\nu\rho\lambda}$ and $L_{\mu\nu\lambda}$ with respect to
$n^\mu$. The tensors $S_{\mu\nu\rho\lambda}$ and $L_{\mu\nu\rho}$ are
related to $\mathcal{S}_{\mu\nu\rho\lambda}$ and
$\mathcal{J}_{\mu\nu\rho}$ by means of the relations
\begin{equation}
S_{\mu\nu\rho\sigma}=\frac{1}{2}(\mathcal{S}_{\mu\nu\rho\sigma}+
\overline{\mathcal{S}}_{\mu\nu\rho\sigma}),\quad
L_{\mu\nu\rho}\equiv\frac{1}{2}(\mathcal{J}_{\mu\nu\rho}+
\overline{\mathcal{J}}_{\mu\nu\rho}),
\end{equation}
and therefore their orthogonal splittings can be calculated once those
of $\mathcal{S}_{\mu\nu\rho\sigma}$ and $\mathcal{J}_{\mu\nu\rho}$ are
known. The orthogonal splitting of $\mathcal{S}_{\mu\nu\rho\sigma}$ is
given in equation (\ref{decompose-ms}) and the orthogonal splitting of
$\mathcal{J}_{\mu\nu\rho}$ can be calculated from equation
(\ref{ms-weylcurrent}) given that we know the orthogonal splitting of
all the quantities which appear in the definition of
$\mathcal{J}_{\mu\nu\rho}$ ---see (\ref{ot-mcalJ}) and
(\ref{ot-mcalZ})-(\ref{ot-mcalP}) in appendix \ref{ot-J} for the
precise formulae. Inserting these orthogonal splittings in
(\ref{q-est}) we get
\begin{eqnarray}
&&n^\nu n^\rho n^\lambda\nabla_\mu B^\mu_{\phantom{\mu}\nu\rho\lambda}=
12\ \mbox{Re}\left[\frac{1}{\sigma-1}\left(\mathcal{E}^\mu Y_{\mu}\mathcal{T}^{\rho\lambda}\overline{\mathcal{T}}_{\rho\lambda}-
2\mathcal{E}^\mu Y^\nu\mathcal{T}_{(\mu}^{\phantom{(\mu}\lambda}\overline{\mathcal{T}}_{\nu)\lambda}\right)\right]\nonumber\\
&&\hspace{5cm}-6\mbox{i}\ \varepsilon_{\mu\nu\rho}\mathcal{T}^{\lambda\nu}
\overline{\mathcal{T}}_{\lambda}^{\phantom{\sigma}\rho}Y\left(\frac{\mathcal{E}^\mu}{\sigma-1}-\frac{\overline{\mathcal{E}}^\mu}
{\overline{\sigma}-1}\right).
\label{step1}
\end{eqnarray}
Now, using the equations (\ref{ots1})-(\ref{ots2}) of appendix
\ref{ot-br} we can deduce the property
$$
B_{\alpha\mu\nu\rho}(\mathcal{E}^\alpha n^\nu n^\rho Y^\mu-\mathcal{E}^\alpha n^\mu n^\nu n^\rho Y)=-\mbox{i}\ \varepsilon_{\mu\nu\alpha}\mathcal{E}^\mu\mathcal{T}^{\rho\nu}\overline{\mathcal{T}}_{\rho}^{\phantom{\rho}\alpha}Y+\mathcal{E}^\mu Y_\mu\mathcal{T}^{\alpha\beta}\overline{\mathcal{T}}_{\alpha\beta}-2\mathcal{E}^\mu Y^\nu\mathcal{T}_{(\nu}^{\phantom{(\nu}\rho}\overline{\mathcal{T}}_{\mu)\rho},
$$
which when combined with (\ref{step1}) yields
\begin{equation}
n^\nu n^\rho n^\sigma\nabla_\mu B^\mu_{\phantom{\mu}\nu\rho\sigma}=
\mbox{Re}\left[\frac{12}{1-\sigma}B_{\mu\nu\alpha\beta}(\mathcal{E}^\mu n^\beta n^\nu n^\alpha Y-\mathcal{E}^\mu n^\alpha n^\beta Y^\nu)\right].
\label{step2}
\end{equation}
Equation (\ref{step2}) is the key to get the required estimate. To
obtain it, let us introduce an orthonormal frame
$E=\{e_0^\mu,e_1^\mu,e_2^\mu,e_3^\mu\}$ with $n^\mu=e_0^\mu$. Since
$\mathcal{E}^\mu$, $Y^\mu$ are spatial vectors they can be written as
a linear combination of the vectors $e_1^\mu,e_2^\mu,e_3^\mu$.
Therefore, point (\ref{second}) of theorem \ref{sproperties} entails the
estimates
\numparts
\begin{eqnarray}
\mbox{Re}\left[\frac{12}{1-\sigma}B_{\mu\nu\alpha\beta}\mathcal{E}^\mu n^\beta n^\nu n^\alpha Y\right]\leq m_1 B_{\mu\nu\alpha\beta}n^\mu n^\nu n^\alpha n^\beta,\\
-\mbox{Re}\left[\frac{12}{1-\sigma}B_{\mu\nu\alpha\beta}
\mathcal{E}^\mu n^\alpha n^\beta Y^\nu\right]\leq m_2
B_{\mu\nu\alpha\beta}n^\mu n^\nu n^\alpha n^\beta,
\end{eqnarray}
\endnumparts
for some scalar continuous functions $m_1$, $m_2$ defined on
$\mathcal{K}$. Using these estimates in equation (\ref{step2}) we
obtain our final estimate $n^\nu n^\rho n^\sigma\nabla_\mu
B^\mu_{\phantom{\mu}\nu\rho\sigma}\leq m_3 B_{\mu\nu\alpha\beta}n^\mu
n^\nu n^\alpha n^\beta$ where $m_3$ is the upper bound of $m_1+m_2$ in
the compact set $\mathcal{K}$ and thus (\ref{gronwall}) can be written
in the form
\begin{equation}
0\leq W'(t)\leq (m_3+3m_0)W(t),\quad \forall t\in [0,t_1).
\end{equation}
Since $W(0)=0$, then an application of Gr\"onwall's lemma
enables us to conclude that $W(t)=0$, $\forall t\in [0,t_1)$ which, according to
the considerations made at the end of section \ref{causal-prop-weyl}
proves the causal propagation.\qed
\begin{remark}\em
We stress that the fact that $\mathcal{S}_{\mu\nu\rho\lambda}$
satisfies a wave equation  and the property
(\ref{covdivmarssimon}) do not, by themselves, suffice to show the
causal propagation of this tensor. If we impose that
$\mathcal{S}_{\mu\nu\rho\lambda}$ and some directional derivative
of this quantity vanish on a spacelike hypersurface
$\mathcal{B}$ then using the wave equation and standard results on
hyperbolic partial differential equations allow us to conclude
that $\mathcal{S}_{\mu\nu\rho\lambda}$ vanishes on a neighbourhood
of $\mathcal{B}$. In doing so, extra conditions needed to be
imposed so that $\mathcal{S}_{\mu\nu\rho\lambda}$ is indeed zero
in $D(\mathcal{B})$.  This approach leads, in principle, to a
weaker result than the one given in theorem \ref{ms-propagation}.
\end{remark}

\smallskip Consider now any
vacuum spacetime admitting a Killing vector $\xi^\mu$ and use $\xi^\mu$
to construct the Mars-Simon tensor $\mathcal{S}_{\mu\nu\rho\lambda}$
and as above, define its real part $S_{\mu\nu\rho\lambda}$ and the
Bel-Robinson tensor $B_{\mu\nu\rho\lambda}$ of
$S_{\mu\nu\rho\lambda}$. For any unit timelike vector $n^\mu$, define
the positive scalar
\begin{equation}
\Phi(\vec{n},\vec{\xi})\equiv B_{\mu\nu\rho\alpha}n^\mu n^\nu n^\rho n^\alpha.
\label{phi}
\end{equation}
One can find an explicit expression for $\Phi(\vec{n},\vec{\xi})$ if we
use the property ---see equation (\ref{ots0}) in
\ref{ot-br}---
\begin{equation}
\fl B_{\mu\nu\rho\alpha}n^\mu n^\nu n^\rho n^\alpha=\mathcal{T}_{\mu\nu}\overline{\mathcal{T}}^{\mu\nu}=
4\mathcal{E}^{\mu\nu}\overline{\mathcal{E}}_{\mu\nu}-24 \mbox{Re}\left[\frac{\mathcal{E}^\mu\mathcal{E}^\nu\mathcal{E}_{\mu\nu}}
{\sigma-1}\right]-12\left|\frac{\mathcal{E}^\mu\mathcal{E}_\mu}{\sigma-1}\right|^2+
36\left|\frac{\mathcal{E}^\mu\overline{\mathcal{E}}_\mu}{\sigma-1}\right|^2,
\label{se-density}
\end{equation}
which shows that $\Phi(\vec{n},\vec{\xi})$ can be expressed in terms
of $\mathcal{E}_{\mu\nu}$, $\mathcal{E}_\mu$ and $\sigma$ exclusively.
When $\Phi(\vec{n},\vec{\xi})$ vanishes on a Cauchy hypersurface, the
causal propagation of $\mathcal{S}_{\mu\nu\rho\lambda}$ entails the
following result.
\begin{theorem}
  Let $(\mathcal{M},g_{\mu\nu})$ be a globally hyperbolic vacuum
  spacetime and suppose that $\mathcal{M}$ admits a Killing vector
  $\xi^\mu$. If there are a Cauchy hypersurface
  $\mathcal{B}\subset\mathcal{M}$ and a real constant $M$ such that
  $\Phi(\vec{N},\vec{\xi})|_{\mathcal{B}}=0$ and
  $\Xi(M,\vec{\xi})|_{\mathcal{B}}=0$ where $\vec{N}$ is the unit
  timelike normal to $\mathcal{B}$, then $\mathcal{M}$ is locally
  isometric to the Kerr spacetime.
\label{positive-f}
\end{theorem}

\noindent
{\it Proof.\hspace{.2cm}}
If $\Phi(\vec{N},\vec{\xi})$ is zero on $\mathcal{B}$
then by (\ref{phi}) $B_{\mu\nu\rho\alpha}N^\mu N^\nu N^\rho N^\alpha|_{\mathcal{B}}=0$.
Therefore, remark \ref{remark:se_density} leads us to
$$
\mathcal{S_{\mu\nu\rho\alpha}}|_{\mathcal{B}}=0.
$$
Using that the Mars-Simon tensor propagates causally, one deduces that $\mathcal{S}_{\mu\nu\rho\alpha}=0$ on $\mathcal{M}$.
In particular, this implies that (\ref{CDXi}) is rendered in the form
\begin{equation}
\nabla_\mu\Xi(M,\vec{\xi})=
\frac{4\sigma_\mu\Xi(M,\vec{\xi})}{\sigma-1},
\label{CDXi0}
\end{equation}
which can be integrated to yield $\Xi=A(1-\sigma)^4$ for some complex
constant $A$. The condition $\Xi(M,\vec{\xi})|_{\mathcal{B}}=0$
implies that $A=0$ and hence $\Xi(M,\vec{\xi})$ vanishes on
$\mathcal{M}$. Theorem \ref{kerr-characterization} can now be applied
and we conclude that $\mathcal{M}$ is locally isometric to the Kerr
spacetime.\qed

\section{Application: characterization of Kerr initial data}
\label{section:application} We explain in this section how to use
previous results to find conditions which ensure that the data
development of a vacuum initial data set is isometric to a subset of
the Kerr spacetime. The basic idea behind is to express
the result of theorem \ref{positive-f} in terms of conditions on
a vacuum initial data set $(\Sigma,h_{ij},K_{ij})$.

\begin{theorem}[\bf Kerr initial data]
Let $(\Sigma,h_{ij},K_{ij})$ be a vacuum initial data set and assume that there exist two scalar fields $\tilde{Y}$, $\tilde{\sigma}$, a vector field
$\tilde{Y}_j$ and a real constant $M$, all defined on $\Sigma$, fulfilling the following conditions
\numparts
\begin{eqnarray}
\fl\mathcal{E}_j\mathcal{E}^j=\frac{1}{4}M^2(1-\tilde{\sigma})^4,\label{e2}\\
\fl 4\mathcal{E}^{ij}\overline{\mathcal{E}}_{ij}-24 \mbox{\em Re}\left[\frac{\mathcal{E}^i\mathcal{E}^j\mathcal{E}_{ij}}
{\tilde{\sigma}-1}\right]-\frac{3}{4}M^4|1-\tilde{\sigma}|^6+
36\left|\frac{\mathcal{E}^i\overline{\mathcal{E}}_i}{\tilde{\sigma}-1}\right|^2=0,
\label{scalar0}\\
\fl D_j\tilde{\sigma}=2\tilde{Y}K_{jl}\tilde{Y}^l-2\ \mbox{i}\ \varepsilon_{jml}
(K_{k}^{\phantom{k}l}\tilde{Y}^k\tilde{Y}^m+\tilde{Y}^m D^l\tilde{Y})+2\tilde{Y}D_j\tilde{Y}+2\tilde{Y}^l D_{[l}\tilde{Y}_{j]},\label{dsigma}\\
\fl D_{(i}\tilde{Y}_{j)}+\tilde{Y} K_{ij}=0, \label{KID1} \\
\fl D_i D_j\tilde{Y}+\mathcal{L}_{\tilde{Y}^l} K_{ij} =\tilde{Y}(r_{ij} + K K_{ij} - 2 K_{il} \updn{K}{l}{j}), \label{KID2}
\end{eqnarray}
\endnumparts
where
\numparts
\begin{eqnarray}
&&\mathcal{E}_j\equiv K_{jk}\tilde{Y}^k+D_j\tilde{Y}-\frac{1}{2}\mbox{\textnormal{i}}\
\varepsilon_{jkl}D^{k}\tilde{Y}^{l},\label{ei}\\
&&
\mathcal{E}_{jk}\equiv\frac{1}{2}(E_{jk}-\mbox{\textnormal{i}}\ B_{jk}).
\end{eqnarray}
\endnumparts
Then the data development $(\mathcal{M},g_{\mu\nu})$ of
$(\Sigma,h_{ij},K_{ij})$ is locally isometric to an open subset of the
Kerr spacetime.
\label{theorem:Kerr_data}
\end{theorem}

\begin{remark}\label{killing_existence}\em The conditions (\ref{KID1}) and (\ref{KID2}) are
  the \emph{Killing initial data} (KID) equations. It is well-known
  that if both a solution $(\tilde{Y},\tilde{Y}_i)$ and the initial
  data set $(\Sigma,h_{ij},K_{ij})$ are \emph{suitably} smooth, then
  the development of the initial data $(\Sigma,h_{ij},K_{ij})$
  possesses a Killing vector such that the pull-back of its
  restriction to $\Sigma$ coincides with the given Killing data ---see
  for example, \cite{COLL,MONCRIEF}, and in particular
  \cite{CHRUSCIEL} for an explicit set of smoothness conditions on the
  Killing data. Alternatively, if the Killing data is
  \emph{transversal}, that is $\tilde{Y}\neq 0$, then one can make use
  of the concept of Killing development to ensure the existence of a
  spacetime with a Killing vector associated to the Killing data under
  question ---see \cite{BEIG}. Again, under suitable regularity
  conditions it can be shown that the Killing development always
  includes the Cauchy development of $(\Sigma,h_{ij},K_{ij})$ ---see
  again \cite{COLL}: the Killing development is unique and maximal
  among the class of spacetimes containing the relevant Cauchy data
  and a Killing vector with non-closed orbits; hence, if the Cauchy
  development has a timelike Killing vector, then it must be a subset
  of the Killing development.
\end{remark}

\noindent
{\it Proof.} Let $\mathcal{M}$ be the data development of the vacuum
initial data set $(\Sigma,h_{ij}, K_{ij})$ and let $\xi^\mu$ be
the Killing vector mentioned in remark \ref{killing_existence}. As usual consider a foliation
$\{\Sigma_t\}$, $t\in I$, of $\mathcal{M}$ with
$\Sigma_0=\phi(\Sigma)$ and let $n^\mu$ be the unit timelike
normal to the leaves of this foliation. Let us adopt the formalism and
notation explained in section \ref{section:mstensor} with $\xi^\mu$ being
the Killing vector. Then the orthogonal splitting of $\xi^\mu$ is
$\xi^\mu=-Y n^\mu+Y^\mu$. As explained in remark \ref{killing_existence} we have the property
\begin{equation}
\phi^*Y=\tilde{Y},\quad \partial_j\phi^\mu Y_\mu=\tilde{Y}_j.
\label{pullback-phi}
\end{equation}
which combined with equations (\ref{ei}) and (\ref{def-mce}) entails
$\partial_j \phi^\mu (\mathcal{E}_\mu)=\mathcal{E}_j$. The orthogonal
splitting of $\sigma_\mu$ is calculated on one hand from equation
(\ref{ernst1}) using the expressions (\ref{decompose-f}),
(\ref{def-mce}) and the property $\xi^\mu=-Y n^\mu+Y^\mu$. The result
is
\begin{equation}
 \sigma_\mu=s_\mu+s n_\mu,
\label{ot-ernst}
\end{equation}
where
\begin{eqnarray}
\fl s\equiv -2K_{\mu\nu}Y^\mu Y^\nu-2Y^\mu D_\mu Y+\mbox{i}\ \varepsilon_{\mu\rho\nu} Y^\mu D^\rho Y^\nu,\nonumber\\
\fl s_\mu\equiv 2 Y K_{\mu\rho}Y^\rho-2\ \mbox{i}\ \varepsilon_{\mu\nu\rho}
(K_{\alpha}^{\phantom{\alpha}\rho}Y^\alpha Y^\nu+Y^\nu D^\rho Y)+2YD_\mu Y+2 Y^\nu D_{[\nu}Y_{\mu]}.\label{def-smu}
\end{eqnarray}
On the other hand, the orthogonal splitting of $\sigma_\mu$ can be calculated from the relation $\sigma_\mu=\nabla_\mu\sigma$ yielding
\begin{equation}
s=-\mathcal{L}_{\vec{n}}\sigma,\quad s_\mu=D_\mu\sigma.
\end{equation}
Combining the previous equation with the pull-back of equation (\ref{def-smu}) under $\phi$ and (\ref{dsigma}) we deduce
\begin{equation}
D_j(\tilde{\sigma})=D_j(\phi^*(\sigma)).
\end{equation}
Upon a suitable choice of the Ernst potential $\sigma$, this equation
can be integrated on $\Sigma$ to give
$\phi^*(\sigma)=\tilde{\sigma}$. Next consider the Mars-Simon tensor
$\mathcal{S}_{\mu\nu\rho\lambda}$ constructed with the Killing vector
$\xi^\mu$ and calculate its orthogonal splitting with respect to
$n^\mu$ ---equation (\ref{decompose-ms}). Using (\ref{tgoestoe}) and our
previous results, we get
\begin{eqnarray}
\fl\phi^*(\Phi(\vec{n},\vec{\xi}))=\phi^*(\mathcal{T}_{\mu\nu}\overline{\mathcal{T}}^{\mu\nu}) \nonumber \\
\fl\hspace{1cm}=\left(
2\overline{\mathcal{E}}_{jk}+\frac{2}{1-\tilde{\sigma}}(
3\mathcal{E}_{j}
\mathcal{E}_{k}-h_{jk}\mathcal{E}_{l}\mathcal{E}^{l})\right)\left(
2\mathcal{E}^{jk}+\frac{2}{1-\overline{\tilde{\sigma}}}(
3\overline{\mathcal{E}}^{j}
\overline{\mathcal{E}}^{k}-h^{jk}\overline{\mathcal{E}}_{m}\overline{\mathcal{E}}^{m})\right).
\end{eqnarray}
Expanding this product and using conditions (\ref{e2})-(\ref{scalar0}) we conclude
\begin{equation}
\Phi(\vec{n},\vec{\xi})|_{\phi(\Sigma)}=0.
\end{equation}
Moreover, equation (\ref{e2}) implies
$\Xi(M,\vec{\xi})|_{\phi(\Sigma)}=0$ ---see
(\ref{ot-f2}). Thus, theorem \ref{positive-f} applies and
therefore we conclude that $(\mathcal{M},g_{\mu\nu})$ must be
locally isometric to an open subset of the Kerr spacetime.\qed

 \subsection{Schwarzschild initial data} \label{Schwarzschild}
Theorem \ref{theorem:Kerr_data} asserts that under certain conditions
the development of a vacuum initial data set is isometric to an open subset
of the Kerr spacetime but nothing is said as to the resulting
development being one of the specializations of the Kerr spacetime:
the Schwarzschild solution. In this section we show the form which
theorem \ref{theorem:Kerr_data} takes in that case. This complements
previous work of the present authors about the same issue
\cite{GARVAL}.

To state our result we need a preliminary lemma whose
proof can be found in
\cite{CarMar08}.
\begin{lemma}
Let $(\Sigma,h_{ij},K_{ij})$ be a vacuum initial data set and assume that there exist a scalar $\tilde{Y}$ and a vector field
$\tilde{Y}_j$, all defined on $\Sigma$, fulfilling the following conditions
\numparts
\begin{eqnarray}
&&\tilde{Y} D_{[i}\tilde{Y}_{j]}+ 2 \tilde{Y}_{[i}D_{j]}\tilde{Y} + 2 \tilde{Y}_{[i}K_{j]l}\tilde{Y}^l =0,\label{kids-1}\\
&&\tilde{Y}_{[i}D_{j}\tilde{Y}_{k]}=0\label{kids-2}.
\end{eqnarray}
\endnumparts
Then there exists an integrable Killing vector $\xi^\mu$ in the data development of $(\Sigma,h_{ij},K_{ij})$.

\label{lemma:static_killing}
\end{lemma}
With the aid of this lemma we can now derive conditions guaranteeing
that the development of a vacuum initial data set is isometric to an
open subset of the Schwarzschild spacetime
\begin{theorem}[{\bf Schwarzschild initial data}] Let
$(\Sigma,h_{ij},K_{ij})$ be a vacuum initial data set and assume
that there exist two scalar fields $\tilde{Y}$, $\tilde{\sigma}$ and a
vector field $\tilde{Y}_j$, all defined on $\Sigma$, fulfilling
the conditions of theorem \ref{theorem:Kerr_data} with
(\ref{KID1})-(\ref{KID2}) replaced by
(\ref{kids-1})-(\ref{kids-2}). Then the data development $\mathcal{M}$
is locally isometric to an open subset of the Schwarzschild spacetime.
\label{theorem:Schwarzschild_data}
\end{theorem}

\noindent
{\it Proof.} The first step in the proof of this theorem is to use
lemma \ref{lemma:static_killing} to show that the data development
$\mathcal{M}$ admits an integrable Killing vector $\xi^\mu$ . The
remaining part of the proof follows the same pattern as that of
theorem \ref{theorem:Kerr_data} and therefore we reach the conclusion
that $\mathcal{M}$ is isometric to an open subset of a specialization
of the Kerr spacetime possessing an integrable Killing vector which,
as is well known, is the Schwarzschild spacetime. \qed

\begin{remark}\em It is worth recalling that the analysis of
\cite{GARVAL} has provided an explicit formula for a \emph{timelike
KID candidate} in terms of concomitants of the initial data
$(h_{ij},K_{ij})$. Hence, a combination of theorem
\ref{theorem:Schwarzschild_data} with the such timelike KID
candidate renders an algorithmic characterization of Schwarzschild
initial data alternative to the one given in \cite{GARVAL}.
\end{remark}

\section{Conclusions} \label{conclusions}

The main result obtained in this article, theorem
\ref{theorem:Kerr_data} provides a characterization of initial data
for the Kerr spacetime by first requiring that a set of overdetermined
elliptic differential equations ---the KID equations--- admit a
solution; and then in turn requiring that certain objects constructed
out of the solutions to the KID equation satisfy very particular
relations. This characterization fails to be algorithmic in two ways.
First, it is in practise a non-trivial problem to verify whether a
certain initial data set admits a KID or not. A pointwise procedure
for doing this has been discussed in \cite{BEICHRSCH}. Alternatively,
one could resort to a non-time symmetric generalization of the
geometric invariant characterizing static time symmetric initial data
sets constructed in \cite{DAI} ---which requires solving a
fourth-order elliptic partial differential equation. In any case
---and this leads to the second way the result fails to be
algorithmic--- even if one could ascertain the existence of a KID on
the initial data, one requires to know the KID in an explicit manner
in order to be able to verify the conditions (\ref{e2}),
(\ref{scalar0}) and (\ref{dsigma}). It is plausible that from an
hypothetical characterization of the Kerr spacetime in terms of, say,
concomitants of the Weyl tensor ---analogous to that for the
Schwarzschild spacetime given in \cite{FerSae}--- one would be able to
construct, following the ideas of \cite{GARVAL}, KID candidates for the
initial data set $(\Sigma,h_{ij},K_{ij})$. If such a KID candidate
were available, then theorem \ref{theorem:Kerr_data} would constitute a
powerful tool to analyse the behaviour of dynamical black hole
spacetimes, both from an analytic and a numerical point of view.

\section*{Acknowledgements}
We would like to thank JMM Senovilla, M Mars, W Simon and CM
Losert-VK for useful remarks on the manuscript.  JAVK is supported
by an EPSRC Advanced Research Fellowship. AGP gratefully
acknowledges financial support from the Spanish ``Ministerio de
Ciencia e Innovaci\'on'' under the postdoctoral grant
EX-2006-0092.

\appendix

\section{Technical details about the calculations}\label{calcdetails}

\subsection{The orthogonal splitting of $\mathcal{J}_{\mu\nu\rho}$}
\label{ot-J}
Given the symmetries of the Weyl current $\mathcal{J}_{\mu\nu\rho}$ it is immediate to realise that its orthogonal splitting must take the form
\begin{equation}
 \mathcal{J}_{\mu\nu\rho}=n_\mu\mathcal{Z}_{\nu\rho}+
n_{[\rho}\mathcal{X}_{\nu]\mu}+n_\mu n_{[\nu}\mathcal{V}_{\rho]}+
\mathcal{P}_{\mu\nu\rho}.
\label{ot-mcalJ}
\end{equation}
It takes a greater effort, however, to find the explicit expression of the spatial tensors
$\mathcal{Z}_{\mu\nu}$, $\mathcal{X}_{\mu\nu}$, $\mathcal{V}_{\mu}$,
$\mathcal{P}_{\mu\nu\rho}$. These expressions are found by inserting into equation (\ref{ms-weylcurrent}) the orthogonal splitting of $\mathcal{S}_{\mu\nu\rho\lambda}$ shown in (\ref{decompose-ms}), the orthogonal splitting of $\mathcal{F}_{\mu\nu}$ shown in (\ref{decompose-f}) and the relation $\xi^\mu=-Y n^\mu+Y^\mu$ . A computer calculation yields the results
%\numparts
\begin{eqnarray}
&&\hspace{-2cm}Z_{\mu\nu}\equiv\frac{4}{1-\sigma}(\mbox{i}\ Y\mathcal{E}^\alpha\mathcal{T}_{[\mu}^{\phantom{[\mu}\rho}\varepsilon_{\nu]\alpha\rho}+3\mathcal{E}_{[\mu}\mathcal{T}_{\nu]\rho}Y^\rho+Y_{[\nu}\mathcal{T}_{\mu]\alpha}\mathcal{E}^\alpha-\mbox{i}\ \varepsilon_{\mu\nu\rho}\mathcal{E}^\alpha\mathcal{T}_\alpha
^{\phantom{\alpha}\rho}Y),\label{ot-mcalZ}\\
&&\hspace{-2cm}\mathcal{X}_{\mu\nu}\equiv
\frac{1}{1-\sigma}\bigg(\mathcal{E}^\alpha\big(4\mbox{i} Y(\mathcal{T}_{(\alpha}^{\phantom{(\alpha}\rho}\varepsilon_{\nu)\mu\rho}-\varepsilon_{\nu\alpha\rho}\mathcal{T}_{\mu}^{\phantom{\mu}\rho})+6\mathcal{T}_{\nu\mu}Y_\alpha-2\mathcal{T}_{\mu\alpha}Y_{\nu}-4\mathcal{T}_{\nu\alpha}Y_\mu+4h_{\nu\mu}\mathcal{T}_{\alpha\rho}Y^\rho \big) \nonumber\\
&&\hspace{1cm}+6\mathcal{E}_\nu\mathcal{T}_{\mu\alpha}Y^\alpha\big) \bigg),\label{ot-mcalX}\\
&&\hspace{-2cm}\mathcal{V}_\mu\equiv\frac{\mathcal{E}^\alpha}{1-\sigma}\left(4\mbox{i}\ Y^\rho(\varepsilon_{\mu\lambda(\rho}\mathcal{T}_{\alpha)}^{\phantom{\alpha)}\lambda}+\varepsilon_{\alpha\rho\lambda}\mathcal{T}_{\mu}^{\phantom{\mu}\lambda})-2Y\mathcal{T}_{\mu\alpha}\right),\label{ot-mcalV}\\
&&\hspace{-2cm}\mathcal{P}_{\mu\nu\rho}\equiv\frac{2}{1-\sigma}\Bigg(2\ \mbox{i}\  \varepsilon{}_{\nu}{}_{\rho}{}_{\lambda} \
\mathcal{E}{}_{\mu} \mathcal{T}{}_{\alpha}{}^{\lambda} Y{}^{\alpha} +
2\mathcal{E}{}_{[\rho} (3Y\mathcal{T}{}_{\nu]\mu} +\mbox{i} \
\varepsilon_{\nu]\mu\lambda}\mathcal{T}{}_{\alpha}{}^{\lambda} \
Y{}^{\alpha}) \nonumber\\
&&\hspace{1cm}+\mathcal{E}{}^{\alpha}\bigg(-2\mbox{i}
\Big(\varepsilon{}_{\rho}{}_{\alpha}{}_{\lambda}\mathcal{T}{}_{[\nu}{}^{\lambda} \
Y{}_{\mu]} +\varepsilon{}_{\nu}{}_{\alpha}{}_{\lambda} \mathcal{T}{}_{[\mu}{}^{\lambda} Y{}_{\rho]}
+2\varepsilon{}_{\mu}{}_{\alpha}{}_{\lambda}\mathcal{T}{}_{[\nu}{}^{\lambda} Y{}_{\rho]}\nonumber \\
&&\hspace{1cm} +Y{}^{\lambda} (\varepsilon{}_{\nu}{}_{\rho}{}_{\alpha} \mathcal{T}{}_{\mu}{}_{\lambda}+\varepsilon{}_{\mu}{}_{\alpha}{}_{[\nu} \mathcal{T}{}_{\rho]}{}_{\lambda})\Big)\nonumber\\
&&\hspace{1cm}+2h{}_{\mu}{}_{[\nu} \Big(\mathcal{T}{}_{\rho]}{}_{\alpha} Y + \mbox{i}\ Y{}^{\lambda} (\varepsilon{}_{\rho]}{}_{\lambda}{}_{\delta} \mathcal{T}{}_{\alpha}{}^{\delta} + \
\mathcal{T}{}_{\rho]}{}^{\delta}\varepsilon{}_{\alpha}{}_{\lambda}{}_{\delta}+ \
\varepsilon{}_{\rho]}{}_{\alpha}{}_{\delta} \mathcal{T}{}_{\lambda}{}^{\delta})\Big)\bigg)\Bigg).\label{ot-mcalP}
\end{eqnarray}
%\endnumparts

\subsection{The orthogonal splitting of $B_{\mu\nu\rho\sigma}$}
\label{ot-br}
In \cite{PARRADO} the orthogonal splitting of the Bel-Robinson tensor was calculated and the different parts resulting from such splitting were studied in detail. The calculations are similar for the Bel-Robinson tensor of a Weyl candidate and therefore we only need to adapt these results for the particular case of $S_{\mu\nu\rho\lambda}$. In this case the orthogonal splitting of $B_{\mu\nu\rho\lambda}$ reads ---see section (5.1) of \cite{PARRADO}---
\begin{equation}
\fl B_{\mu\nu\rho\lambda}=Wn_\mu n_\nu n_\rho n_\lambda+4P_{(\mu}n_\nu n_\rho n_{\lambda)}+6t_{(\mu\nu}n_\rho n_{\lambda)}
+4Q_{(\mu\nu\rho}n_{\lambda)}+t_{\mu\nu\rho\lambda},
\label{br-decomposition}
\end{equation}
where the different parts of this splitting are given by
%\numparts
\begin{eqnarray}
&&W\equiv e_{\alpha\beta}e^{\alpha\beta}+b_{\alpha\beta}b^{\alpha\beta},\\
&&P_\mu\equiv 2b_{\alpha}^{\phantom{\alpha} \lambda}e_{\rho\lambda}\varepsilon_\mu^{\phantom{\mu}\alpha\rho},\\
&&t_{\mu\nu}\equiv W h_{\mu\nu}-2(b_\mu^{\phantom{\mu}\rho}b_{\nu\rho}+e_\mu^{\phantom{\mu}\rho}e_{\nu\rho}),\\
&&Q_{\mu\nu\rho}\equiv h_{\mu\nu}P_\rho -2
\left(b_{\nu\alpha}e_{\mu\beta}+b_{\mu\alpha}
e_{\nu\beta}\right) \varepsilon_{\rho}^{\phantom{\rho}\alpha\beta},\label{br-parts}\\
&&t_{\mu\nu\rho\lambda}^{{    }}\equiv 4(b_{{\mu\nu}}^{{  }}
b_{{\rho\lambda}}^{{  }}+e_{{\mu\nu}}^{{  }} e_{{\rho\lambda}}^{{  }})
-h_{{\rho\lambda}} t_{{\mu\nu}}^{{  }}+2h_{{\nu(\lambda}}
t_{{\rho)\mu}}+2h_{{\mu(\lambda}}t_{{\rho)\nu}}-h_{{\mu\nu}}^{{  }} t_{{\rho\lambda}}+\nonumber\\
&&\hspace{2cm}+W(h_{\mu\nu}h_{\rho\lambda}-2h_{\mu(\rho} h_{\lambda)\nu}),
\end{eqnarray}
%\endnumparts
where $e_{\mu\nu}$ and $b_{\mu\nu}$ are respectively the {\em
electric} and {\em magnetic} parts of the Weyl candidate
$S_{\mu\nu\lambda\rho}$. From (\ref{br-decomposition}) and
(\ref{br-parts}) we get
%\numparts
\begin{eqnarray}
&&\fl B_{\mu\nu\rho\lambda}n^\mu n^\nu n^\rho n^\lambda=\mathcal{T}_{\rho\lambda}\overline{\mathcal{T}}^{\rho\lambda}
\label{ots0}\\
&&\fl B_{\mu\nu\rho\lambda}n^\nu n^\rho n^\lambda=
-n_\mu\mathcal{T}^{\rho\lambda}\overline{\mathcal{T}}_{\rho\lambda}
+\mbox{i}\ \varepsilon_{\mu\nu\rho}\mathcal{T}^{\lambda\nu}\overline{\mathcal{T}}_{\lambda}^{\phantom{\lambda}\rho}.\label{ots1}\\
&&\fl B_{\mu\nu\rho\lambda}n^\rho n^\lambda=-2\mathcal{T}_{(\mu}^{\phantom{(\mu}\rho}\mathcal{T}_{\nu)\rho}+
(n_{\mu}n_{\nu}+h_{\mu\nu})\mathcal{T}^{\rho\lambda}\overline{\mathcal{T}}_{\rho\lambda}-2\mbox{i}\ \mathcal{T}^{\alpha\rho}\overline{\mathcal{T}}_{\alpha}^
{\phantom{\alpha}\lambda}\varepsilon_{\rho\lambda(\mu}n_{\nu)},\label{ots2}
\end{eqnarray}
%\endnumparts
where the relations
$$
e_{\mu\nu}=\mathcal{T}_{\mu\nu}+\overline{\mathcal{T}}_{\mu\nu},\
b_{\mu\nu}=\mbox{i}\ (\mathcal{T}_{\mu\nu}-\overline{\mathcal{T}}_{\mu\nu}),
$$
need to be used.

\end{document}